\documentclass[prl,twocolumn,floatfix,superscriptaddress,longbibliography,notitlepage]{revtex4-1}

\usepackage[latin9]{inputenc}
\usepackage{color}
\usepackage{amsmath}
\usepackage{amssymb}
\usepackage{stmaryrd}
\usepackage{graphicx}
\usepackage{romannum}
\usepackage{physics}
\usepackage{tcolorbox}

\usepackage[unicode=true,
 bookmarks=true,bookmarksnumbered=false,bookmarksopen=false,
 breaklinks=false,pdfborder={0 0 1},backref=false,colorlinks=true]
 {hyperref}
\hypersetup{
 linkcolor=magenta, urlcolor=blue, citecolor=blue, pdfstartview={FitH},unicode=true}
 
\makeatletter
\@ifundefined{textcolor}{}
{%
 \definecolor{BLACK}{gray}{0}
 \definecolor{WHITE}{gray}{1}
 \definecolor{RED}{rgb}{1,0,0}
 \definecolor{GREEN}{rgb}{0,1,0}
 \definecolor{BLUE}{rgb}{0,0,1}
 \definecolor{CYAN}{cmyk}{1,0,0,0}
 \definecolor{MAGENTA}{cmyk}{0,1,0,0}
 \definecolor{YELLOW}{cmyk}{0,0,1,0}
}

\newcommand{\doublewidetilde}[1]{{%
  \mathpalette\double@widetilde{#1}%
}}
\newcommand{\double@widetilde}[2]{%
  \sbox\z@{$\m@th#1\widetilde{#2}$}%
  \ht\z@=.9\ht\z@
  \widetilde{\box\z@}%

}

\newcommand{\ryd}{\texttt{Rydberg}}
\newcommand{\noryd}{\texttt{No Rydberg}}

\definecolor{darkgreen}{rgb}{0.0, 0.6, 0.13}

\usepackage{amsfonts}\usepackage{tabularx}\usepackage{dcolumn}\usepackage{bm}\usepackage{graphicx}\usepackage{epstopdf}

\hypersetup{urlcolor=blue}

\usepackage{tensor}
\usepackage{braket}

\usepackage[capitalise,compress]{cleveref}

\makeatletter
\DeclareFontFamily{OMX}{MnSymbolE}{}
\DeclareSymbolFont{MnLargeSymbols}{OMX}{MnSymbolE}{m}{n}
\SetSymbolFont{MnLargeSymbols}{bold}{OMX}{MnSymbolE}{b}{n}
\DeclareFontShape{OMX}{MnSymbolE}{m}{n}{
    <-6>  MnSymbolE5
   <6-7>  MnSymbolE6
   <7-8>  MnSymbolE7
   <8-9>  MnSymbolE8
   <9-10> MnSymbolE9
  <10-12> MnSymbolE10
  <12->   MnSymbolE12
}{}
\DeclareFontShape{OMX}{MnSymbolE}{b}{n}{
    <-6>  MnSymbolE-Bold5
   <6-7>  MnSymbolE-Bold6
   <7-8>  MnSymbolE-Bold7
   <8-9>  MnSymbolE-Bold8
   <9-10> MnSymbolE-Bold9
  <10-12> MnSymbolE-Bold10
  <12->   MnSymbolE-Bold12
}{}

\let\llangle\@undefined
\let\rrangle\@undefined
\DeclareMathDelimiter{\llangle}{\mathopen}%
                     {MnLargeSymbols}{'164}{MnLargeSymbols}{'164}
\DeclareMathDelimiter{\rrangle}{\mathclose}%
                     {MnLargeSymbols}{'171}{MnLargeSymbols}{'171}
\makeatother

\begin{document}
\title{Quantum Non-Demolition Photon Counting in a 2d Rydberg Atom Array}

\author{ Christopher Fechisin}\thanks{Correspondence to: fechisin@umd.edu.}
  \affiliation{Joint Quantum Institute, NIST/University of Maryland, College Park, MD, 20742, USA}
    \affiliation{Joint Center for Quantum Information and Computer Science, NIST/University of Maryland, College Park, MD, 20742, USA}
  
\author{ Kunal Sharma}
  \affiliation{Joint Center for Quantum Information and Computer Science, NIST/University of Maryland, College Park, MD, 20742, USA}
 \affiliation{IBM~Quantum,~IBM~T.J.~Watson~Research~Center,~Yorktown~Heights,~NY~10598,~USA}
 
 \author{ Przemyslaw Bienias}
  \affiliation{Joint Quantum Institute, NIST/University of Maryland, College Park, MD, 20742, USA}
  \affiliation{Joint Center for Quantum Information and Computer Science, NIST/University of Maryland, College Park, MD, 20742, USA}

 \author{ Steven L. Rolston}
  \affiliation{Joint Quantum Institute, NIST/University of Maryland, College Park, MD, 20742, USA}

 \author{ J. V. Porto}
  \affiliation{Joint Quantum Institute, NIST/University of Maryland, College Park, MD, 20742, USA}

 \author{ Michael J. Gullans}
  \affiliation{Joint Center for Quantum Information and Computer Science, NIST/University of Maryland, College Park, MD, 20742, USA}
 
  \author{ Alexey V. Gorshkov}
 \affiliation{Joint Quantum Institute, NIST/University of Maryland, College Park, MD, 20742, USA}
 \affiliation{Joint Center for Quantum Information and Computer Science, NIST/University of Maryland, College Park, MD, 20742, USA}
 
\date{\today}
\begin{abstract}

Rydberg arrays merge the collective behavior of ordered atomic arrays with the controllability and optical nonlinearities of Rydberg systems, resulting in a powerful platform for realizing photonic many-body physics. As an application of this platform, we propose a protocol for quantum non-demolition (QND) photon counting. Our protocol involves photon storage in the Rydberg array, an observation phase consisting of a series of Rabi flops to a Rydberg state and measurements, and retrieval of the stored photons. The Rabi frequency experiences a $\sqrt{n}$ collective enhancement, where $n$ is the number of photons stored in the array. Projectively measuring the presence or absence of a Rydberg excitation after oscillating for some time is thus a weak measurement of photon number. We demonstrate that the photon counting protocol can be used to distill Fock states from arbitrary pure or mixed initial states and to perform photonic state discrimination. We confirm that the protocol still works in the presence of experimentally realistic noise.

\end{abstract}
\maketitle

\pagenumbering{arabic}
Quantum non-demolition photon counting is a key paradigm in quantum optics with direct applications in quantum information processing and quantum networking. Photon counting is most simply performed by capturing photons with a detector which converts each photon to an electrical signal \cite{becker_advanced_2005,oconnor_time-correlated_2012}, but this process destroys the quantum state of the photons. Methods of QND photon counting circumvent this issue in various ways, preserving the quantum state of the photons after the measurement \cite{liang_quantum_2014,brune_quantum_1990,guerlin_progressive_2007,dotsenko_quantum_2009,sayrin_real-time_2011,liu_quantum_2020,holland_quantum_1991,reiserer_nondestructive_2013,schuster_resolving_2007,johnson_quantum_2010,malz_nondestructive_2020,grimsmo_quantum_2021}. This is ideal for applications like quantum networking and state preparation which make further use of the photonic state after the measurement.

Rather than directly measuring the photon number, it is sometimes preferable to make a series of weak measurements via projective measurement of an alternative observable which is more readily accessible experimentally. Many weak measurements performed sequentially can progressively collapse the system into an eigenstate of the photon number with high fidelity \cite{brune_quantum_1990, dotsenko_quantum_2009, guerlin_progressive_2007, haroche_manipulating_2009, peaudecerf_adaptive_2014, sayrin_real-time_2011}. So long as the series of weak measurements does not destroy the quantum coherence of the photons, it serves as an effective QND measurement of the photon number. 

Methods for QND photon counting have been studied in a wide variety of experimental platforms, including cold atomic gases \cite{liang_quantum_2014}, microwave cavities \cite{brune_quantum_1990,guerlin_progressive_2007,dotsenko_quantum_2009,sayrin_real-time_2011,liu_quantum_2020}, optical cavities \cite{holland_quantum_1991,reiserer_nondestructive_2013}, superconducting circuits \cite{schuster_resolving_2007,johnson_quantum_2010,royer_itinerant_2018}, waveguides \cite{shahmoon_strongly_2011,malz_nondestructive_2020}, and nonlinear metamaterials \cite{grimsmo_quantum_2021}.
These proposals each carry drawbacks and advantages relative to one another, and can be ideal in different situations. Protocols which encode photon number in a phase, for instance, are well-suited for resolving small photon numbers, but are only well-defined within a single period of the phase \cite{guerlin_progressive_2007}. Some proposals approach the task of non-destructively counting itinerant photons \cite{reiserer_nondestructive_2013,besse_single-shot_2018,kono_quantum_2018,royer_itinerant_2018,shahmoon_strongly_2011,malz_nondestructive_2020,grimsmo_quantum_2021}, while others require confinement to a cavity \cite{liang_quantum_2014,brune_quantum_1990,holland_quantum_1991,guerlin_progressive_2007,dotsenko_quantum_2009,sayrin_real-time_2011,liu_quantum_2020,schuster_resolving_2007,johnson_quantum_2010}. The protocol which we study in this work has no fundamental limitation in discerning large or small photon numbers and is able to count itinerant photons by storing them in a Rydberg array in free space. 

\begin{figure}
\centering\includegraphics[width=0.4\textwidth]{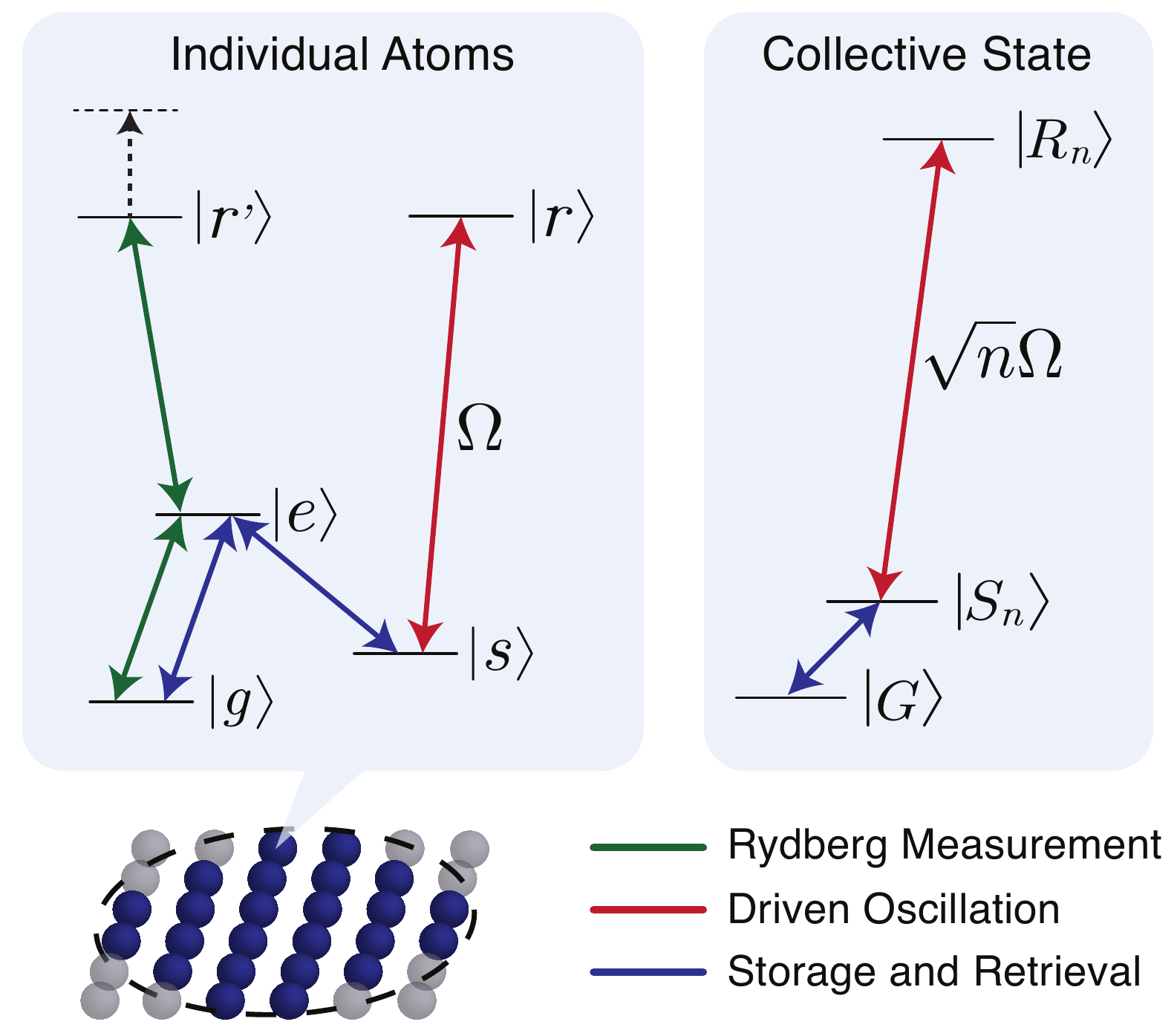}
  \caption{The level structure of each atom in the array, as well as the collective states of the array. The dashed line indicates the shift in the Rydberg level $\ket{r'}$ due to the presence of a Rydberg excitation $\ket{r}$. This disrupts the EIT condition by pushing the $\ket{e}$ to $\ket{r'}$ transition far off resonance, making the system fully reflective in the presence of an $\ket{r}$ excitation.
  }
  \label{fig:levels}
\end{figure}

Rydberg atoms, due to their intrinsic controllability and strong dipole-dipole and van der Waals interactions, have become a prototypical system for facilitating interactions between photons \cite{gorshkov_photon-photon_2011,bienias_scattering_2014,zeuthen_correlated_2017,bienias_exotic_2020,ornelas-huerta_tunable_2021} and, empowered by the development of Rydberg arrays, simulating many-body physics \cite{maghrebi_fractional_2015,omran_generation_2019,celi_emerging_2020,samajdar_complex_2020,semeghini_probing_2021,kalinowski_bulk_2022,bluvstein_quantum_2022}. The quantum optical properties of disordered ensembles of Rydberg atoms are well-known \cite{sun_analysis_2018,ripka_room-temperature_2018,tiarks_photonphoton_2019,ornelas-huerta_-demand_2020}, but the quantum optics of ordered Rydberg arrays is less well-understood and has been the subject of much recent theoretical \cite{bekenstein_quantum_2020,zhang_photon-photon_2021,moreno-cardoner_quantum_2021,pedersen_quantum_2022-1} and experimental \cite{srakaew_subwavelength_2022} work. Ordered atomic arrays have been shown to exhibit emergent behaviors arising from cooperative interactions, such as acting as near-perfect mirrors \cite{bettles_enhanced_2016,shahmoon_cooperative_2017,rui_subradiant_2020}, emitting light in fixed, geometrically determined directions \cite{moreno-cardoner_quantum_2021}, and---most importantly for this work---storing light with significantly increased fidelity \cite{manzoni_optimization_2018}. Rydberg arrays combine these collective phenomena with strong optical nonlinearities, making for a powerful platform for the realization of photonic many-body physics \cite{wei_generation_2021} and, as we study in this work, QND photon counting.

\begin{figure*}
  \centering\includegraphics[width=\textwidth]{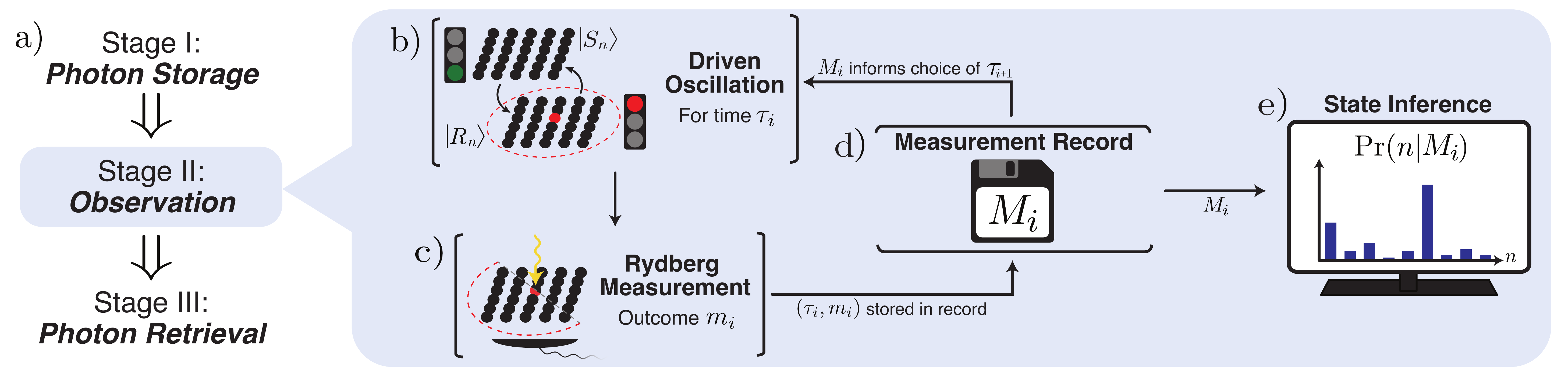}
  \caption{\textbf{a)} The three stages of the protocol.  Stage II is the focus of this work. \textbf{b)} Driven oscillation between the collective states $\ket{S_n}$ and $\ket{R_n}$, where the red atom is in the Rydberg state and the dashed line depicts the blockade radius. The traffic light reflects the fact that the array permits the probe light to pass only when there is no Rydberg excitation present. This is because the array is tuned to EIT, which is disrupted by a Rydberg excitation. \textbf{c)-d)} Projective measurement in the $\{\ryd{}, \noryd{}\}$ basis, the results of which are stored in the measurement record $M_i$. The two sides of the array indicate the possible measurement outcomes. The measurement is performed by shining weak classical light on the array, indicated by the yellow arrow. \textbf{e)} The measurement record is used to infer the state of the system. Here we schematically depict an intermediate inference based on a small measurement record $M_i$ which has not yet converged to a Fock state.
  }
  \label{fig2}
\end{figure*}

Our protocol consists of three stages: (I) storing a photonic pulse in the array, (II) measuring the number of photons $n$ contained therein, and (III) retrieving the stored pulse. The photon number is progressively pinned down through a series of observation cycles, each of which consists of a partial Rabi flop to a Rydberg spin wave state followed by a direct measurement of the presence of a Rydberg excitation. The frequency of the Rabi flop is enhanced by a factor of $\sqrt{n}$ from the single-atom Rabi frequency, making it possible to discern arbitrary $n$. The outcome of the Rydberg measurement along with the known evolution time thus serves as a weak measurement of photon number $n$.

Our main result is that, in the absence of dephasing noise during the observation stage, we can perform QND detection of $n$ photons with fidelity limited only by storage and retrieval error in time $t_{\gamma=0}\sim \sqrt{n}/\Omega$, where $\Omega$ is the single atom Rabi frequency and $\gamma$ is the dephasing rate. In the presence of dephasing noise, our detection is no longer completely non-demolition and takes time $t_{\gamma>0}\sim\gamma n/\Omega^2$. We discuss later precisely how destructive the protocol is in the presence of noise.

{\it The physical system.---}We consider an ordered two-dimensional array of atoms with subwavelength spacing, where each atom has the level structure shown in \cref{fig:levels}. Lowercase letters label single-atom states, while uppercase letters label collective states of the many-body system. In particular, $\ket{g}$ is a ground state, $\ket{e}$ is an excited state, $\ket{s}$ is a metastable shelving state, and $\ket{r}$ and $\ket{r'}$ are Rydberg states. In a later section, we will propose particular levels in Yb to realize these states. 

$\ket{G}$ is the many-body ground state in which all atoms are in the state $\ket{g}$. $\ket{S_n}$ is the symmetrized collective state with $n$ excitations of individual atoms to $\ket{s}$: $|S_n\rangle = \frac{1}{\sqrt{_NC_n}}\sum_{i_j=1}^N \hat\sigma_{sg}^{(i_1)}\cdots\hat\sigma_{sg}^{(i_n)}|G\rangle$, where $\hat \sigma_{ij}:= \ketbra{i}{j}$. Similarly, $\ket{R_n}$ is the symmetrized spin wave state with $n-1$ atoms excited to $\ket{s}$ and one atom excited to the Rydberg state $\ket{r}$: $\ket{R_n}=\frac{1}{\sqrt{n }}\sum_{i=1}^N \hat\sigma_{rs}^{(i)}|S_n\rangle$. We assume that the entire array is within a blockade radius, so that further excitations to $\ket{r}$ are forbidden by the blockade.

{\it The protocol.---}In Stage I, an initial photonic state couples to the $\ket{g}-\ket{e}$ transition and is stored in the array using an auxiliary classical control field acting on the $\ket{s}-\ket{e}$ transition \cite{manzoni_optimization_2018}. 
The initial photonic state is sent through a beamsplitter so that it is normally incident upon the array symmetrically from both sides, as is necessary for optimal storage efficiency \cite{manzoni_optimization_2018}. We denote the photonic state as a superposition of number states $\ket{n}$, $\ket{\psi_\text{ph}}=\sum_{i=1}^N c_n\ket{n}$. $N$ is the number of atoms in the array and therefore the upper bound on the number of excitations which can be stored in the array. Photon storage is performed on the $\ket{g}-\ket{e}-\ket{s}$ $\Lambda$-subsystem as described in \cite{manzoni_optimization_2018}. Storage maps the state of the array from $\ket{G}$ to $\sum_{n}c_n|S_n\rangle$, where the amplitudes $c_n$ are inherited from $\ket{\psi_\text{ph}}$. Each photon is therefore stored as an excitation from $\ket{g}$ to $\ket{s}$. 

Stage II consists of two operations on the collective state: (1) Rabi flops between the states $\ket{s}$ and $\ket{r}$ and (2) projective measurement of the presence of a Rydberg excitation (see \cref{fig2}). To drive the collective oscillation, we couple $|s_i\rangle$ to the Rydberg state $|r_i\rangle$ via the rotating frame Hamiltonian
$
\hat H=\Omega\sum_{i=1}^N\hat\sigma_{rs}^{(i)}+\text{h.c.},
$
where $\Omega$ is the Rabi frequency of the transition and the sum is over all atoms in the array. This induces a coupling between the collective states $\ket{S_n}$ and $\ket{R_n}$ which depends explicitly on the number of stored photons $n$: $\langle S_n|\hat H|R_n\rangle=\sqrt{n}\Omega=: \Omega_n$. The objective of this stage is to indirectly and progressively measure the photon number $n$ by directly and repeatedly measuring the presence of a Rydberg excitation after evolution under $\hat H$ for some known time. Each observation cycle consists of a driven oscillation for time $\tau_i$ and a projective measurement of the collective state yielding the measurement outcome $m_i\in\{\ryd{},\noryd{}\}$, building over time a measurement record through cycle $T$ denoted $M_T=\{(\tau_i,m_i)\}_{i=1,\ldots,T}$. In \cref{ap:adaptive}, we discuss how to choose times $\{\tau_i\}$ to expedite convergence.

 Our measurements project the array into the subspace with or without a Rydberg excitation in $\ket{r}$, conditioned on the outcome of the measurement. We are able to efficiently perform such a measurement by tuning to EIT and applying a weak classical probe light. EIT is applied to the $\ket{g}-\ket{e}-\ket{r'}$ subsystem, with the $\ket{s}-\ket{r}$ subsystem acting as a `switch'. Thus in the absence of a Rydberg excitation, the EIT condition is satisfied and the array is transmissive, but in the presence of a Rydberg excitation the EIT condition is disrupted and the array is once again near-perfectly reflective \cite{moreno-cardoner_quantum_2021}. This measurement takes finite time which is limited by the width of the EIT transparency window.
 We note that a similar mechanism was used in \cite{honer_artificial_2011,murray_photon_2018} for performing photon subtraction with atomic clouds. 

Stage III consists of retrieval of the stored photons. Assuming that the measured photon number was $n$, the array must be in the state $\ket{S_n}$ to retrieve the photons. If the preceding measurement instead projected the state to $\ket{R_n}$, we can simply drive for time $\tau=\pi/2\Omega_n$ and arrive at $\ket{S_n}$, so long as the measurement has converged to $n$. Retrieval can then be performed as in \cite{manzoni_optimization_2018}. The final photonic state comes out symmetrically from both sides of the array and can be combined onto a single path using a beamsplitter. 

{\it Measurement dynamics.---}The focus of our analysis is Stage II of the protocol. The measurement dynamics of the system during this stage determine the protocol's capabilities and effectiveness. This stage also
raises the question of how to best choose driving times $\tau_i$.

The dynamics under driving are described by the following Hamiltonian written in the basis of collective excitations:~$
\hat H_\text{coll}=\sum_{n}\Omega_n\left(\ketbra{R_n}{S_n}+\ketbra{S_n}{R_n}\right).
$~We begin in the state~$|\psi(0)\rangle=\sum_n c_n|S_n\rangle$ with amplitudes $c_n$ inherited from the stored light pulse. Driving under~$\hat H_\text{coll}$ for time $\tau$ yields~$
|\psi(\tau)\rangle=\sum_n c_n(\cos(\Omega_n\tau)|S_n\rangle-i\sin(\Omega_n\tau)|R_n\rangle).
$ 
Performing the Rydberg measurement as described in the previous section results in two possible outcomes:~$|\psi_S\rangle = \frac{1}{p_{S}}\sum_{n} c_n\cos{(\Omega_n\tau)}|S_n\rangle$ or
         $|\psi_R\rangle = \frac{1}{p_{R}}\sum_{n} c_n\sin{(\Omega_n\tau)}|R_n\rangle$,
with probabilities $p_{S}=\sum_n|c_n\cos{(\Omega_n\tau)}|^2$ and $p_{R}=\sum_n|c_n\sin{(\Omega_n\tau)}|^2$. The amplitudes $c_n$ are therefore updated by a factor proportional to $\cos{(\Omega_n\tau)}$ or $\sin{(\Omega_n\tau)}$ as a result of the measurement, so that $c^{(1)}_n\sim c_n\sin{(\Omega_n\tau)}$ or $c^{(1)}_n\sim c_n\cos{(\Omega_n\tau)}$, depending on the measurement outcome. This reflects the partial information learned about the photon number from a single measurement of the collective state. 

After each measurement, the system is once again in a state of the form $|\psi\rangle=\sum_n c^{(i)}_n|X_n\rangle$, where $X\in\{S,R\}$. Further iterations of unitary evolution and measurement will continue to update the amplitudes, so that after the $i^\text{th}$ observational cycle the amplitudes are given by $\{c_n^{(i)}\}$. For an arbitrary pure or mixed initial state (see next section), the state converges to a single photon number state as the protocol iterates, analogous to the progressive state collapse observed in \cite{guerlin_progressive_2007}, with the outcome always the distillation of a single photon number state.
 
{\it Generalization to mixed states.---}The analysis of the preceding sections readily generalizes to mixed intial photonic states. The dynamics in this case are governed by the master equation $\dot\rho(t)=-i[\hat H_\text{coll},\rho]$, under which the populations of $\rho$ evolve independently of the coherences in the Fock basis. This implies that any two density matrices $\rho$ and $\rho'$ which satisfy $\rho_{ii}(0)=\rho'_{ii}(0)$ will also satisfy $\rho_{ii}(t)=\rho'_{ii}(t)$. In particular, for any mixed state $\rho(0)$ there exists some pure state $\rho'(0)$ such that $\rho_{ii}(0)=\rho'_{ii}(0)$ and therefore $\rho_{ii}(t)=\rho'_{ii}(t)$. This has several important implications for the performance of the protocol which are discussed in \cref{ap:mixed}.

{\it Initial state inference.---}An important application of QND photon counting is state inference---gleaning information about the initial photonic state from the measurement record $M_T=\{(\tau_i,m_i)\}_{i=1,\ldots,T}$. Consider the set of distributions of photon number states with maximum photon number $N$, which we denote $\{P_\alpha\}_N$. Each distribution $P_\alpha=(p_0,\ldots,p_N)\in\{P_\alpha\}_N$ describes a class of potential initial photonic states which share the same populations. Note that we cannot discern within these classes, as the protocol is insensitive to coherences between number states. Bayes' Theorem yields the probability that the initial state was described by $P_\alpha$, conditioned on the measurement record $M_T$:
$
\text{Pr}(P_\alpha|M_T)=\frac{\text{Pr}(M_T|P_\alpha)\text{Pr}(P_\alpha)}{\sum_{\alpha'}\text{Pr}(M_T|P_{\alpha'})\text{Pr}(P_{\alpha'})}
$
, where $\text{Pr}(P_\alpha)$ is a prior over the set of initial distributions. $\text{Pr}(M_T|P_\alpha)$ the quantity which we must compute to find $\text{Pr}(P_\alpha|M_T)$. Because different number states are not mixed through the protocol, this expression can be written
$
\text{Pr}(M_T|P_\alpha)=\sum_n p_n \text{Pr}(M_T|n),
$
where $\text{Pr}(M_T|n)$ is the probability of observing the measurement record $M_T$ given initial state $|S_n\rangle$. This is readily given by
$\text{Pr}(M_T|n)=\prod_{(\tau_i,m_i)\in M_T}\text{Pr}(m_i;t_i|n)$, where
\begin{align}\label{eq:prod_sines_cosines}
\text{Pr}(m_i;t_i|n)=\left\{\begin{array}{cl}
         \cos{(\Omega_n\tau_i)}^2, & m_i=m_{i-1}\\
         \sin{(\Omega_n\tau_i)}^2, & m_i\neq m_{i-1}
    \end{array}\right.
\end{align}
and $m_0=\noryd{}$.

{\it Performance in the presence of noise.---}Dephasing is a significant form of noise in Rydberg array experiments \cite{burgers_controlling_2022}, therefore we study the protocol in the presence of dephasing noise and propose potential modifications which improve performance under high noise. In particular, we study the evolution of the density matrix $\rho(t)$ of the system under the master equation
$
    \dot\rho(t)=-i[\hat H,\rho]+\gamma\left(\sum_i \hat n_i\rho \hat n_i-\frac{1}{2}\{\hat n_i,\rho\}\right),
$
where the jump operators $\ketbra{r_i}{r_i}=: \hat n_i=\hat n_i^\dagger=\hat n_i^\dagger \hat n_i$ represent dephasing on the state $\ket{r}$ at the $i$th atom. This model is very similar to that studied in \cite{honer_artificial_2011}, and reproduces several features found therein.

This noise model can be understood as a continuous description of the process in which the environment measures where the Rydberg excitation is located within the spin wave, thereby removing the measured site from the coherent and symmetric superposition. Given enough time, such processes will remove all sites from the spin wave, resulting in a state which is entirely decohered. This is reflected in an exponential decay in the amplitude of the driven oscillation between $\ket{S_n}$ and $\ket{R_n}$. 

\begin{figure*}
    \centering
    \includegraphics[width = 0.95\textwidth]{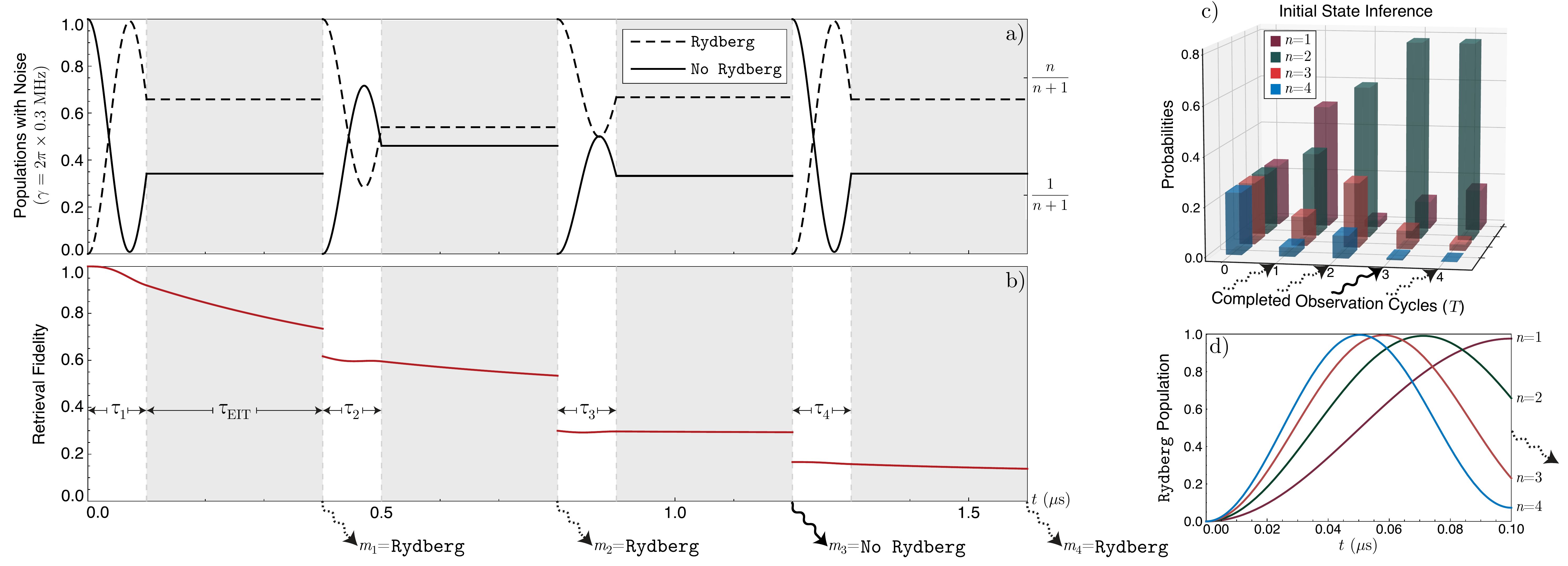}
    \caption{An example experiment in which the state inference converges to $n=2$. We have set $\Omega=2\pi\times 2.5$ MHz and $\gamma=2\pi\times .3$ MHz. We take the detection time to be $\tau_\text{EIT} = .3$ $\mu$s, the inverse of the EIT transparency window  $1/\tau_\text{EIT}=\Omega^2/\Gamma_c \approx 2\pi\times0.5$ MHz, where $\Gamma_c = 2\pi\times 12$ MHz is the subradiant linewidth. \cite{zhang_photon-photon_2021}. 
    \textbf{a)} The populations in the \ryd{} and \noryd{} sectors as a function of time. This is the signal upon which the photon number is imprinted. The gray boxes correspond to Rydberg measurements during which $\Omega$ is turned off but dephasing persists. The dashing of the wavy arrow indicates the result of the Rydberg measurement. \textbf{b)}  The fidelity of retrieving the full 2-photon pulse. Notice that the retrieval fidelity decays during the driven oscillation as well as the Rydberg measurement. \textbf{c)} The state inference as a function of observation cycle. The initial inference $(T=0)$ is uniform because we have chosen a uniform prior over $n=1$--$4$. \textbf{d)} The \ryd{} populations of various $n$ during the first observation cycle. Notice that the state inference in c) simply reflects the values of these probabilities at $t=0.10$ $\mu$s as this is the only information available after the first observation cycle.}
    \label{fig:efficient_new}
\end{figure*}

In the presence of noise, it is necessary to modify the likelihood $\text{Pr}(M_T|n)$ appearing in the state inference to include dephasing. This reflects the fact that we are seeking to match observations with a different underlying signal. We denote the likelihood in the presence of noise by $\text{Pr}(M_T|n,\gamma)$. As in the absence of noise, this quantity is a product: $\text{Pr}(M_T|n,\gamma)=\prod_{(m_i,t_i)\in M_T}\Pr(m_i;t_i|n,\gamma).$
We can calculate $\Pr(m_i;t_i|n,\gamma)$ numerically (see \cref{fig:efficient_new}), working in a truncated Hilbert space which, due to site permutation symmetry, is of dimension linear in $n$. The details of this approach can be found in \cref{ap:numerics}. This can be done in practice because the dephasing rate $\gamma$ can be measured for a particular experimental realization of the protocol and therefore enters as a known parameter.

{\it Modifications in the presence of noise.---}The previous section describes the simplest extension of our protocol to a noisy environment, and we may improve performance in the presence of noise by modifying the protocol more dramatically. 

Within our noise model, the evolution of the system is particularly sensitive to the measurement outcome. When a Rydberg excitation is detected, the array is projected into the \ryd{} sector, and further time evolution continues the approach of the system to its long-time steady state. In contrast, when no Rydberg excitation is detected, the decoherence between $\ket{r}$ and $\ket{s}$ is effectively `reset'. Going too long without measuring \noryd{} therefore leads to a weakened signal, which slows Fock state distillation and state inference. This slow-down can be avoided by ejecting the Rydberg atom when a Rydberg excitation is detected, returning the system to the \noryd{} sector of the Hilbert space with one fewer atom. In doing so, we prevent the detection time from scaling as $\gamma n/\Omega^2$ rather than $\sqrt{n}/\Omega$, at the cost of losing a number of photons throughout the protocol.
 
In the regime where $\gamma$ is large, oscillations rapidly decay, and we are no longer able to use frequency as a probe of photon number. However, the photon number is also imprinted on the steady state values of the populations in the \ryd{} and \noryd{} sectors, which are $\frac{n}{n+1}$ and $\frac{1}{n+1}$, respectively. These values emerge because the master equation drives the density matrix to the maximally mixed state in the Fock basis, which has $n$ \ryd{} states for each \noryd{} state, which was also described in \cite{honer_artificial_2011}.
The steady-state populations are captured by $\text{Pr}(M_T|n,\gamma)$, so no modifications to the state inference procedure are necessary to probe this signals. However, this signal is much weaker than the frequency signal, with detection time scaling as $n^3/\gamma$. The detection is also no longer QND as the atomic state will have largely decohered. 

{\it Detection Time.---}In this section, we discuss how the time necessary to resolve the photon number scales with the parameters of the problem. To proceed quantitatively, we consider the Fisher information $F_n(t)=\mathbb{E}\left[\left(\frac{\partial}{\partial n}\log f(t;n)\right)^2\right]$, where $f(t;n)$ stands for the various signals---noiseless oscillation frequency, noisy oscillation frequency, and steady-state populations---from which we can infer the parameter $n$. In our case, we have two possible measurement outcomes \ryd{} and \noryd{}, so that $F_n(t)=f_{m_i\neq m_{i-1}}(t;n)\left(\frac{\partial}{\partial n}\log f_{m_i\neq m_{i-1}}(t;n)\right)^2+f_{m_i= m_{i-1}}(t;n)\left(\frac{\partial}{\partial n}\log f_{m_i= m_{i-1}}(t;n)\right)^2$. The subscript denotes whether the probability is that of measuring the same or opposite outcome as that previously measured. We define the detection time $t_*$ via $F_n(t_*)=1$, as this is when we have sufficient information to discern $n$ from $n+1$.

In the absence of noise, we have $f_{m_i= m_{i-1}}(t;n)=\cos^2(\sqrt{n}\Omega t)$ and $f_{m_i\neq m_{i-1}}(t;n)=\sin^2(\sqrt{n}\Omega t)$. Using these values, we find that $F_n(t)=t^2\Omega^2/n$ so that $t_*=\sqrt{n}/\Omega$ in the absence of noise.

In order to approximate the convergence time in the presence of noise, we assume that we can approximate the signal as noiseless, so long as we perform the measurement in a short time relative to the inverse of the dephasing rate $T=1/\gamma$. In practice, we would choose a time $\alpha T$ where $\alpha<1$ is a positive constant, but here we simply use $T$ as we are only interested in scaling behaviors. Then for a single measurement we once again find $F_n(T)=T^2\Omega^2/n$. However, in time $t$, we are able to make $t/T$ measurements. To account for this, we multiply by a shot noise factor $t/T=t\gamma$. We then find that $F_n(t)=t\Omega^2/\gamma n$ so that $t_*=\gamma n/\Omega^2$ in the presence of noise.

Finally, we consider the case in which $\gamma$ is large and we only have access to the signal imprinted on the steady-state populations, rather than on the oscillation frequency. In this case we would like to wait long enough to approximately approach the steady-state populations before making each measurement. The characteristic time scale is still $T$, though in this case we would choose a time $\alpha T$ where $\alpha >1$ in practice. Once again we simply take this time to be $T$ to capture the scaling behavior. The signal in this case is $f_S(t;n)=1/(n+1)$ and $f_R(t;n)=n/(n+1)$, where the subscripts correspond to \noryd{} and \ryd{} because the signal is a function of the current measurement outcome only, not the previous. In this case, we find that $F_n(T)=1/n(n+1)^2$. Multiplying by the shot noise factor $t\gamma$, we have $F_n(t)=t\gamma/n(n+1)^2$ and $t_*=n (1 + n)^2/\gamma\sim n^3/\gamma$.

{\it Experimental considerations.---}We now discuss two potential implementations of our protocol in ytterbium atoms trapped in an array of optical tweezers. Yb is a particularly attractive atomic platform because it has a long wavelength telecom transition (studied in \cite{covey_telecom-band_2019} and proposed for use in Rydberg arrays in \cite{bekenstein_quantum_2020}), and because trapping of its Rydberg states in the same tweezer as the ground state has been recently demonstrated, extending trap lifetimes to $\sim50-100$ $\mu$s \cite{wilson_trapping_2022}.

One of the implementations which we study uses $^{171}$Yb, and the other $^{174}$Yb. These implementations only differ in the states assigned to $\ket{g}$ and $\ket{s}$. In $^{171}$Yb, $\ket{g}$ and $\ket{s}$ are assigned to two hyperfine $^3$P$_0$ states: $\ket{g}:=\ket{^3\text{P}_0,\downarrow}$ and $\ket{s}:=\ket{^3\text{P}_0,\uparrow}$, where the arrow represents the nuclear spin. In $^{174}$Yb, they are assigned to two members of the $^3$P$_J$ triplet: $\ket{g}:=\ket{^3\text{P}_0}$ and $\ket{s}:=\ket{^3\text{P}_2}$.

In both implementations, we assign $\ket{e}:=\ket{^3\text{D}_1}$, making use of the $^3$P$_0\leftrightarrow ^3$D$_1$ telecom transition referenced above, which is approximately 1.4 $\mu$m. This wavelength, $\lambda_{eg}$, sets an upper bound on the array spacing $d$. Choosing levels such that $\lambda_{eg}$ is large therefore allows for larger tweezer spacing without sacrificing efficiency. The Rydberg levels are assigned as $\ket{r}:=\ket{6sns\text{ }^3\text{S}_1}$ and $\ket{r'}:=\ket{m'pn's\text{ }^3\text{P}_J}$, where $m'$, $n'$ and $J$ are chosen to maximize the matrix element from $\ket{s}$ while avoiding dipole-dipole population exchange between $\ket{r}$ and $\ket{r'}$.

There is a potential complication to using hyperfine states in $^{171}$Yb for photon storage. As discussed in \cite{manzoni_optimization_2018}, hyperfine structure opens the possibility of atoms decaying to multiple ground states, thus weakening the subradiance which underpins much of the valuable physics hosted by ordered arrays. Maximal subradiance can be (at least partially) restored by applying large Zeeman shifts, or potentially by engineering patterns of entanglement in the many-body states which suppress emmission which is not subradiant \cite{asenjo-garcia_optical_2019}.

{\it Discussion and outlook.---}In summary, we have presented a noise-resilient protocol to non-destructively count arbitrary numbers of free space photons using an ordered array of Rydberg atoms. We have shown that this protocol is effective in distilling arbitrary pure and mixed initial states to a single Fock state and in performing initial state inference. Within our analysis, the overall efficiency of the protocol is limited by the storage and retrieval efficiencies as well as dephasing during the observation stage. There is strong theoretical evidence that the storage and retrieval errors are significantly suppressed in our array geometry \cite{manzoni_optimization_2018}, and the dephasing errors may in some cases be suppressed by further cooling the system.

In principle, the experiment could instead be performed using a disordered atomic ensemble without significant changes to the protocol. However, disordered ensembles do not enjoy the strongly enhanced storage and retrieval efficiencies of ordered arrays. A disordered ensemble with a relatively high optical depth per blockade radius (ODb) of 10-12 has an optimal combined storage and retrieval efficiency of $\sim 50\%$ \cite{gorshkov_photon_2007}, compared to an array of $16=4\times 4$ atoms with an optimal combined storage and retrieval efficiency above $98\%$ \cite{manzoni_optimization_2018}. It could however be beneficial to consider the experiment taking place in a cavity instead of in free space. This
could further enhance the storage and retrieval efficiency of the array or even make disordered ensembles a more viable experimental platform.

Another interesting extension would be to consider photonic density matrix tomography. Because displaced Fock states are tomographically complete \cite{wallentowitz_unbalanced_1996,banaszek_direct_1996,nehra_state-independent_2019}, one would only need to introduce the additional operation of displacements in phase space, given by $D(\alpha)=e^{\alpha \hat a^\dagger-\alpha^* \hat a}$ where $\hat a^\dagger = \sum_n \sqrt{n+1} \ketbra{S_{n+1}}{S_n}$ and $\hat a = \sum_n \sqrt{n} \ketbra{S_{n}}{S_{n+1}}$. This can be easily implemented  on the stored excitations in our protocol via the couplings
$
        \hat H_x = \Omega \sum_j \hat\sigma_{s,g}^{(j)}+ \hat\sigma_{g,s}^{(j)}\approx\Omega\sqrt{N-n}\left(\hat a^\dagger +\hat a\right)$ and $
        \hat H_y = i\Omega \sum_j \hat\sigma_{s,g}^{(j)}- \hat\sigma_{g,s}^{(j)}\approx i\Omega\sqrt{N-n}\left(\hat a^\dagger-\hat a\right),$
where the second equality in each holds in the limit $N-n\gg 1$. Any $D(\alpha)$ can be implemented by successively applying $\hat H_x$ for time $t_x$ and $\hat H_y$ for time $t_y$, with $\alpha=\Omega\sqrt{N-n}(it_x-t_y)$.

\begin{acknowledgments} 
We thank Sharoon Austin, Patrick Banner, and Deniz Kurdak for helpful discussions. CF thanks the Joint Quantum Institute at the University of Maryland for support through a JQI fellowship. This work was supported by DARPA SAVaNT ADVENT, AFOSR, AFOSR MURI, the DoE ASCR Quantum Testbed Pathfinder program (award No.~DE-SC0019040), DoE QSA, NSF QLCI (award No.~OMA-2120757), DoE ASCR Accelerated Research in Quantum Computing program (award No.~DE-SC0020312), NSF PFCQC program, ARO MURI, and the U.S. Army Research Laboratory's Maryland ARL Quantum Partnership (W911NF-19-2-0181).
 
\end{acknowledgments}

\bibliographystyle{bibstyle}
\bibliography{Rydberg_QND_References_Better,10.3.22}

\begin{thebibliography}{57}%
\makeatletter
\providecommand \@ifxundefined [1]{%
 \@ifx{#1\undefined}
}%
\providecommand \@ifnum [1]{%
 \ifnum #1\expandafter \@firstoftwo
 \else \expandafter \@secondoftwo
 \fi
}%
\providecommand \@ifx [1]{%
 \ifx #1\expandafter \@firstoftwo
 \else \expandafter \@secondoftwo
 \fi
}%
\providecommand \natexlab [1]{#1}%
\providecommand \enquote  [1]{``#1''}%
\providecommand \bibnamefont  [1]{#1}%
\providecommand \bibfnamefont [1]{#1}%
\providecommand \citenamefont [1]{#1}%
\providecommand \href@noop [0]{\@secondoftwo}%
\providecommand \href [0]{\begingroup \@sanitize@url \@href}%
\providecommand \@href[1]{\@@startlink{#1}\@@href}%
\providecommand \@@href[1]{\endgroup#1\@@endlink}%
\providecommand \@sanitize@url [0]{\catcode `\\12\catcode `\$12\catcode
  `\&12\catcode `\#12\catcode `\^12\catcode `\_12\catcode `\%12\relax}%
\providecommand \@@startlink[1]{}%
\providecommand \@@endlink[0]{}%
\providecommand \url  [0]{\begingroup\@sanitize@url \@url }%
\providecommand \@url [1]{\endgroup\@href {#1}{\urlprefix }}%
\providecommand \urlprefix  [0]{URL }%
\providecommand \Eprint [0]{\href }%
\providecommand \doibase [0]{http://dx.doi.org/}%
\providecommand \selectlanguage [0]{\@gobble}%
\providecommand \bibinfo  [0]{\@secondoftwo}%
\providecommand \bibfield  [0]{\@secondoftwo}%
\providecommand \translation [1]{[#1]}%
\providecommand \BibitemOpen [0]{}%
\providecommand \bibitemStop [0]{}%
\providecommand \bibitemNoStop [0]{.\EOS\space}%
\providecommand \EOS [0]{\spacefactor3000\relax}%
\providecommand \BibitemShut  [1]{\csname bibitem#1\endcsname}%
\let\auto@bib@innerbib\@empty
\bibitem [{\citenamefont {Becker}(2005)}]{becker_advanced_2005}%
  \BibitemOpen
  \bibfield  {author} {\bibinfo {author} {\bibfnamefont {W.}~\bibnamefont
  {Becker}},\ }\href@noop {} {\emph {\bibinfo {title} {Advanced
  {{Time-Correlated Single Photon Counting Techniques}}}}}\ (\bibinfo
  {publisher} {{Springer Science \& Business Media}},\ \bibinfo {year}
  {2005})\BibitemShut {NoStop}%
\bibitem [{\citenamefont {O'Connor}(2012)}]{oconnor_time-correlated_2012}%
  \BibitemOpen
  \bibfield  {author} {\bibinfo {author} {\bibfnamefont {D.}~\bibnamefont
  {O'Connor}},\ }\href@noop {} {\emph {\bibinfo {title} {Time-Correlated Single
  Photon Counting}}}\ (\bibinfo  {publisher} {{Academic Press}},\ \bibinfo
  {year} {2012})\BibitemShut {NoStop}%
\bibitem [{\citenamefont {Liang}\ \emph {et~al.}(2014)\citenamefont {Liang},
  \citenamefont {Lin}, \citenamefont {Hao}, \citenamefont {Niu},\ and\
  \citenamefont {Gong}}]{liang_quantum_2014}%
  \BibitemOpen
  \bibfield  {author} {\bibinfo {author} {\bibfnamefont {L.}~\bibnamefont
  {Liang}}, \bibinfo {author} {\bibfnamefont {G.~W.}\ \bibnamefont {Lin}},
  \bibinfo {author} {\bibfnamefont {Y.~M.}\ \bibnamefont {Hao}}, \bibinfo
  {author} {\bibfnamefont {Y.~P.}\ \bibnamefont {Niu}}, \ and\ \bibinfo
  {author} {\bibfnamefont {S.~Q.}\ \bibnamefont {Gong}},\ }\bibinfo {title}
  {\emph {Quantum Nondemolition Measurement of Small Photon Numbers Using
  Stored Light}},\ \href {\doibase 10.1103/PhysRevA.90.055801} {\bibfield
  {journal} {\bibinfo  {journal} {Physical Review A}\ }\textbf {\bibinfo
  {volume} {90}},\ \bibinfo {pages} {055801} (\bibinfo {year}
  {2014})}\BibitemShut {NoStop}%
\bibitem [{\citenamefont {Brune}\ \emph {et~al.}(1990)\citenamefont {Brune},
  \citenamefont {Haroche}, \citenamefont {Lefevre}, \citenamefont {Raimond},\
  and\ \citenamefont {Zagury}}]{brune_quantum_1990}%
  \BibitemOpen
  \bibfield  {author} {\bibinfo {author} {\bibfnamefont {M.}~\bibnamefont
  {Brune}}, \bibinfo {author} {\bibfnamefont {S.}~\bibnamefont {Haroche}},
  \bibinfo {author} {\bibfnamefont {V.}~\bibnamefont {Lefevre}}, \bibinfo
  {author} {\bibfnamefont {J.~M.}\ \bibnamefont {Raimond}}, \ and\ \bibinfo
  {author} {\bibfnamefont {N.}~\bibnamefont {Zagury}},\ }\bibinfo {title}
  {\emph {Quantum Nondemolition Measurement of Small Photon Numbers by
  {{Rydberg-atom}} Phase-Sensitive Detection}},\ \href {\doibase
  10.1103/PhysRevLett.65.976} {\bibfield  {journal} {\bibinfo  {journal}
  {Physical Review Letters}\ }\textbf {\bibinfo {volume} {65}},\ \bibinfo
  {pages} {976} (\bibinfo {year} {1990})}\BibitemShut {NoStop}%
\bibitem [{\citenamefont {Guerlin}\ \emph {et~al.}(2007)\citenamefont
  {Guerlin}, \citenamefont {Bernu}, \citenamefont {Del{\'e}glise},
  \citenamefont {Sayrin}, \citenamefont {Gleyzes}, \citenamefont {Kuhr},
  \citenamefont {Brune}, \citenamefont {Raimond},\ and\ \citenamefont
  {Haroche}}]{guerlin_progressive_2007}%
  \BibitemOpen
  \bibfield  {author} {\bibinfo {author} {\bibfnamefont {C.}~\bibnamefont
  {Guerlin}}, \bibinfo {author} {\bibfnamefont {J.}~\bibnamefont {Bernu}},
  \bibinfo {author} {\bibfnamefont {S.}~\bibnamefont {Del{\'e}glise}}, \bibinfo
  {author} {\bibfnamefont {C.}~\bibnamefont {Sayrin}}, \bibinfo {author}
  {\bibfnamefont {S.}~\bibnamefont {Gleyzes}}, \bibinfo {author} {\bibfnamefont
  {S.}~\bibnamefont {Kuhr}}, \bibinfo {author} {\bibfnamefont {M.}~\bibnamefont
  {Brune}}, \bibinfo {author} {\bibfnamefont {J.-M.}\ \bibnamefont {Raimond}},
  \ and\ \bibinfo {author} {\bibfnamefont {S.}~\bibnamefont {Haroche}},\
  }\bibinfo {title} {\emph {Progressive Field-State Collapse and Quantum
  Non-Demolition Photon Counting}},\ \href {\doibase 10.1038/nature06057}
  {\bibfield  {journal} {\bibinfo  {journal} {Nature}\ }\textbf {\bibinfo
  {volume} {448}},\ \bibinfo {pages} {889} (\bibinfo {year}
  {2007})}\BibitemShut {NoStop}%
\bibitem [{\citenamefont {Dotsenko}\ \emph {et~al.}(2009)\citenamefont
  {Dotsenko}, \citenamefont {Mirrahimi}, \citenamefont {Brune}, \citenamefont
  {Haroche}, \citenamefont {Raimond},\ and\ \citenamefont
  {Rouchon}}]{dotsenko_quantum_2009}%
  \BibitemOpen
  \bibfield  {author} {\bibinfo {author} {\bibfnamefont {I.}~\bibnamefont
  {Dotsenko}}, \bibinfo {author} {\bibfnamefont {M.}~\bibnamefont {Mirrahimi}},
  \bibinfo {author} {\bibfnamefont {M.}~\bibnamefont {Brune}}, \bibinfo
  {author} {\bibfnamefont {S.}~\bibnamefont {Haroche}}, \bibinfo {author}
  {\bibfnamefont {J.-M.}\ \bibnamefont {Raimond}}, \ and\ \bibinfo {author}
  {\bibfnamefont {P.}~\bibnamefont {Rouchon}},\ }\bibinfo {title} {\emph
  {Quantum Feedback by Discrete Quantum Nondemolition Measurements: {{Towards}}
  on-Demand Generation of Photon-Number States}},\ \href {\doibase
  10.1103/PhysRevA.80.013805} {\bibfield  {journal} {\bibinfo  {journal}
  {Physical Review A}\ }\textbf {\bibinfo {volume} {80}},\ \bibinfo {pages}
  {013805} (\bibinfo {year} {2009})}\BibitemShut {NoStop}%
\bibitem [{\citenamefont {Sayrin}\ \emph {et~al.}(2011)\citenamefont {Sayrin},
  \citenamefont {Dotsenko}, \citenamefont {Zhou}, \citenamefont {Peaudecerf},
  \citenamefont {Rybarczyk}, \citenamefont {Gleyzes}, \citenamefont {Rouchon},
  \citenamefont {Mirrahimi}, \citenamefont {Amini}, \citenamefont {Brune},
  \citenamefont {Raimond},\ and\ \citenamefont
  {Haroche}}]{sayrin_real-time_2011}%
  \BibitemOpen
  \bibfield  {author} {\bibinfo {author} {\bibfnamefont {C.}~\bibnamefont
  {Sayrin}}, \bibinfo {author} {\bibfnamefont {I.}~\bibnamefont {Dotsenko}},
  \bibinfo {author} {\bibfnamefont {X.}~\bibnamefont {Zhou}}, \bibinfo {author}
  {\bibfnamefont {B.}~\bibnamefont {Peaudecerf}}, \bibinfo {author}
  {\bibfnamefont {T.}~\bibnamefont {Rybarczyk}}, \bibinfo {author}
  {\bibfnamefont {S.}~\bibnamefont {Gleyzes}}, \bibinfo {author} {\bibfnamefont
  {P.}~\bibnamefont {Rouchon}}, \bibinfo {author} {\bibfnamefont
  {M.}~\bibnamefont {Mirrahimi}}, \bibinfo {author} {\bibfnamefont
  {H.}~\bibnamefont {Amini}}, \bibinfo {author} {\bibfnamefont
  {M.}~\bibnamefont {Brune}}, \bibinfo {author} {\bibfnamefont {J.-M.}\
  \bibnamefont {Raimond}}, \ and\ \bibinfo {author} {\bibfnamefont
  {S.}~\bibnamefont {Haroche}},\ }\bibinfo {title} {\emph {Real-Time Quantum
  Feedback Prepares and Stabilizes Photon Number States}},\ \href {\doibase
  10.1038/nature10376} {\bibfield  {journal} {\bibinfo  {journal} {Nature}\
  }\textbf {\bibinfo {volume} {477}},\ \bibinfo {pages} {73} (\bibinfo {year}
  {2011})}\BibitemShut {NoStop}%
\bibitem [{\citenamefont {Liu}\ \emph {et~al.}(2020)\citenamefont {Liu},
  \citenamefont {Chen},\ and\ \citenamefont {Segal}}]{liu_quantum_2020}%
  \BibitemOpen
  \bibfield  {author} {\bibinfo {author} {\bibfnamefont {J.}~\bibnamefont
  {Liu}}, \bibinfo {author} {\bibfnamefont {H.-T.}\ \bibnamefont {Chen}}, \
  and\ \bibinfo {author} {\bibfnamefont {D.}~\bibnamefont {Segal}},\ }\bibinfo
  {title} {\emph {Quantum Nondemolition Photon Counting with a Hybrid
  Electromechanical Probe}},\ \href {\doibase 10.1103/PhysRevA.102.061501}
  {\bibfield  {journal} {\bibinfo  {journal} {Physical Review A}\ }\textbf
  {\bibinfo {volume} {102}},\ \bibinfo {pages} {061501} (\bibinfo {year}
  {2020})}\BibitemShut {NoStop}%
\bibitem [{\citenamefont {Holland}\ \emph {et~al.}(1991)\citenamefont
  {Holland}, \citenamefont {Walls},\ and\ \citenamefont
  {Zoller}}]{holland_quantum_1991}%
  \BibitemOpen
  \bibfield  {author} {\bibinfo {author} {\bibfnamefont {M.~J.}\ \bibnamefont
  {Holland}}, \bibinfo {author} {\bibfnamefont {D.~F.}\ \bibnamefont {Walls}},
  \ and\ \bibinfo {author} {\bibfnamefont {P.}~\bibnamefont {Zoller}},\
  }\bibinfo {title} {\emph {Quantum Nondemolition Measurements of Photon Number
  by Atomic Beam Deflection}},\ \href {\doibase 10.1103/PhysRevLett.67.1716}
  {\bibfield  {journal} {\bibinfo  {journal} {Physical Review Letters}\
  }\textbf {\bibinfo {volume} {67}},\ \bibinfo {pages} {1716} (\bibinfo {year}
  {1991})}\BibitemShut {NoStop}%
\bibitem [{\citenamefont {Reiserer}\ \emph {et~al.}(2013)\citenamefont
  {Reiserer}, \citenamefont {Ritter},\ and\ \citenamefont
  {Rempe}}]{reiserer_nondestructive_2013}%
  \BibitemOpen
  \bibfield  {author} {\bibinfo {author} {\bibfnamefont {A.}~\bibnamefont
  {Reiserer}}, \bibinfo {author} {\bibfnamefont {S.}~\bibnamefont {Ritter}}, \
  and\ \bibinfo {author} {\bibfnamefont {G.}~\bibnamefont {Rempe}},\ }\bibinfo
  {title} {\emph {Nondestructive {{Detection}} of an {{Optical Photon}}}},\
  \href {\doibase 10.1126/science.1246164} {\bibfield  {journal} {\bibinfo
  {journal} {Science}\ }\textbf {\bibinfo {volume} {342}},\ \bibinfo {pages}
  {1349} (\bibinfo {year} {2013})},\ \Eprint {http://arxiv.org/abs/1311.3625}
  {arXiv:1311.3625}\BibitemShut {NoStop}%
\bibitem [{\citenamefont {Schuster}\ \emph {et~al.}(2007)\citenamefont
  {Schuster}, \citenamefont {Houck}, \citenamefont {Schreier}, \citenamefont
  {Wallraff}, \citenamefont {Gambetta}, \citenamefont {Blais}, \citenamefont
  {Frunzio}, \citenamefont {Majer}, \citenamefont {Johnson}, \citenamefont
  {Devoret}, \citenamefont {Girvin},\ and\ \citenamefont
  {Schoelkopf}}]{schuster_resolving_2007}%
  \BibitemOpen
  \bibfield  {author} {\bibinfo {author} {\bibfnamefont {D.~I.}\ \bibnamefont
  {Schuster}}, \bibinfo {author} {\bibfnamefont {A.~A.}\ \bibnamefont {Houck}},
  \bibinfo {author} {\bibfnamefont {J.~A.}\ \bibnamefont {Schreier}}, \bibinfo
  {author} {\bibfnamefont {A.}~\bibnamefont {Wallraff}}, \bibinfo {author}
  {\bibfnamefont {J.~M.}\ \bibnamefont {Gambetta}}, \bibinfo {author}
  {\bibfnamefont {A.}~\bibnamefont {Blais}}, \bibinfo {author} {\bibfnamefont
  {L.}~\bibnamefont {Frunzio}}, \bibinfo {author} {\bibfnamefont
  {J.}~\bibnamefont {Majer}}, \bibinfo {author} {\bibfnamefont
  {B.}~\bibnamefont {Johnson}}, \bibinfo {author} {\bibfnamefont {M.~H.}\
  \bibnamefont {Devoret}}, \bibinfo {author} {\bibfnamefont {S.~M.}\
  \bibnamefont {Girvin}}, \ and\ \bibinfo {author} {\bibfnamefont {R.~J.}\
  \bibnamefont {Schoelkopf}},\ }\bibinfo {title} {\emph {Resolving Photon
  Number States in a Superconducting Circuit}},\ \href {\doibase
  10.1038/nature05461} {\bibfield  {journal} {\bibinfo  {journal} {Nature}\
  }\textbf {\bibinfo {volume} {445}},\ \bibinfo {pages} {515} (\bibinfo {year}
  {2007})}\BibitemShut {NoStop}%
\bibitem [{\citenamefont {Johnson}\ \emph {et~al.}(2010)\citenamefont
  {Johnson}, \citenamefont {Reed}, \citenamefont {Houck}, \citenamefont
  {Schuster}, \citenamefont {Bishop}, \citenamefont {Ginossar}, \citenamefont
  {Gambetta}, \citenamefont {DiCarlo}, \citenamefont {Frunzio}, \citenamefont
  {Girvin},\ and\ \citenamefont {Schoelkopf}}]{johnson_quantum_2010}%
  \BibitemOpen
  \bibfield  {author} {\bibinfo {author} {\bibfnamefont {B.~R.}\ \bibnamefont
  {Johnson}}, \bibinfo {author} {\bibfnamefont {M.~D.}\ \bibnamefont {Reed}},
  \bibinfo {author} {\bibfnamefont {A.~A.}\ \bibnamefont {Houck}}, \bibinfo
  {author} {\bibfnamefont {D.~I.}\ \bibnamefont {Schuster}}, \bibinfo {author}
  {\bibfnamefont {L.~S.}\ \bibnamefont {Bishop}}, \bibinfo {author}
  {\bibfnamefont {E.}~\bibnamefont {Ginossar}}, \bibinfo {author}
  {\bibfnamefont {J.~M.}\ \bibnamefont {Gambetta}}, \bibinfo {author}
  {\bibfnamefont {L.}~\bibnamefont {DiCarlo}}, \bibinfo {author} {\bibfnamefont
  {L.}~\bibnamefont {Frunzio}}, \bibinfo {author} {\bibfnamefont {S.~M.}\
  \bibnamefont {Girvin}}, \ and\ \bibinfo {author} {\bibfnamefont {R.~J.}\
  \bibnamefont {Schoelkopf}},\ }\bibinfo {title} {\emph {Quantum Non-Demolition
  Detection of Single Microwave Photons in a Circuit}},\ \href {\doibase
  10.1038/nphys1710} {\bibfield  {journal} {\bibinfo  {journal} {Nature
  Physics}\ }\textbf {\bibinfo {volume} {6}},\ \bibinfo {pages} {663} (\bibinfo
  {year} {2010})}\BibitemShut {NoStop}%
\bibitem [{\citenamefont {Malz}\ and\ \citenamefont
  {Cirac}(2020)}]{malz_nondestructive_2020}%
  \BibitemOpen
  \bibfield  {author} {\bibinfo {author} {\bibfnamefont {D.}~\bibnamefont
  {Malz}}\ and\ \bibinfo {author} {\bibfnamefont {J.~I.}\ \bibnamefont
  {Cirac}},\ }\bibinfo {title} {\emph {Nondestructive Photon Counting in
  Waveguide {{QED}}}},\ \href {\doibase 10.1103/PhysRevResearch.2.033091}
  {\bibfield  {journal} {\bibinfo  {journal} {Physical Review Research}\
  }\textbf {\bibinfo {volume} {2}},\ \bibinfo {pages} {033091} (\bibinfo {year}
  {2020})}\BibitemShut {NoStop}%
\bibitem [{\citenamefont {Grimsmo}\ \emph {et~al.}(2021)\citenamefont
  {Grimsmo}, \citenamefont {Royer}, \citenamefont {Kreikebaum}, \citenamefont
  {Ye}, \citenamefont {O'Brien}, \citenamefont {Siddiqi},\ and\ \citenamefont
  {Blais}}]{grimsmo_quantum_2021}%
  \BibitemOpen
  \bibfield  {author} {\bibinfo {author} {\bibfnamefont {A.~L.}\ \bibnamefont
  {Grimsmo}}, \bibinfo {author} {\bibfnamefont {B.}~\bibnamefont {Royer}},
  \bibinfo {author} {\bibfnamefont {J.~M.}\ \bibnamefont {Kreikebaum}},
  \bibinfo {author} {\bibfnamefont {Y.}~\bibnamefont {Ye}}, \bibinfo {author}
  {\bibfnamefont {K.}~\bibnamefont {O'Brien}}, \bibinfo {author} {\bibfnamefont
  {I.}~\bibnamefont {Siddiqi}}, \ and\ \bibinfo {author} {\bibfnamefont
  {A.}~\bibnamefont {Blais}},\ }\bibinfo {title} {\emph {Quantum
  {{Metamaterial}} for {{Broadband Detection}} of {{Single Microwave
  Photons}}}},\ \href {\doibase 10.1103/PhysRevApplied.15.034074} {\bibfield
  {journal} {\bibinfo  {journal} {Physical Review Applied}\ }\textbf {\bibinfo
  {volume} {15}},\ \bibinfo {pages} {034074} (\bibinfo {year}
  {2021})}\BibitemShut {NoStop}%
\bibitem [{\citenamefont {Haroche}\ \emph {et~al.}(2009)\citenamefont
  {Haroche}, \citenamefont {Dotsenko}, \citenamefont {Del{\'e}glise},
  \citenamefont {Sayrin}, \citenamefont {Zhou}, \citenamefont {Gleyzes},
  \citenamefont {Guerlin}, \citenamefont {Kuhr}, \citenamefont {Brune},\ and\
  \citenamefont {Raimond}}]{haroche_manipulating_2009}%
  \BibitemOpen
  \bibfield  {author} {\bibinfo {author} {\bibfnamefont {S.}~\bibnamefont
  {Haroche}}, \bibinfo {author} {\bibfnamefont {I.}~\bibnamefont {Dotsenko}},
  \bibinfo {author} {\bibfnamefont {S.}~\bibnamefont {Del{\'e}glise}}, \bibinfo
  {author} {\bibfnamefont {C.}~\bibnamefont {Sayrin}}, \bibinfo {author}
  {\bibfnamefont {X.}~\bibnamefont {Zhou}}, \bibinfo {author} {\bibfnamefont
  {S.}~\bibnamefont {Gleyzes}}, \bibinfo {author} {\bibfnamefont
  {C.}~\bibnamefont {Guerlin}}, \bibinfo {author} {\bibfnamefont
  {S.}~\bibnamefont {Kuhr}}, \bibinfo {author} {\bibfnamefont {M.}~\bibnamefont
  {Brune}}, \ and\ \bibinfo {author} {\bibfnamefont {J.-M.}\ \bibnamefont
  {Raimond}},\ }\bibinfo {title} {\emph {Manipulating and Probing Microwave
  Fields in a Cavity by Quantum Non-Demolition Photon Counting}},\ \href
  {\doibase 10.1088/0031-8949/2009/T137/014014} {\bibfield  {journal} {\bibinfo
   {journal} {Physica Scripta}\ }\textbf {\bibinfo {volume} {T137}},\ \bibinfo
  {pages} {014014} (\bibinfo {year} {2009})}\BibitemShut {NoStop}%
\bibitem [{\citenamefont {Peaudecerf}\ \emph {et~al.}(2014)\citenamefont
  {Peaudecerf}, \citenamefont {Rybarczyk}, \citenamefont {Gerlich},
  \citenamefont {Gleyzes}, \citenamefont {Raimond}, \citenamefont {Haroche},
  \citenamefont {Dotsenko},\ and\ \citenamefont
  {Brune}}]{peaudecerf_adaptive_2014}%
  \BibitemOpen
  \bibfield  {author} {\bibinfo {author} {\bibfnamefont {B.}~\bibnamefont
  {Peaudecerf}}, \bibinfo {author} {\bibfnamefont {T.}~\bibnamefont
  {Rybarczyk}}, \bibinfo {author} {\bibfnamefont {S.}~\bibnamefont {Gerlich}},
  \bibinfo {author} {\bibfnamefont {S.}~\bibnamefont {Gleyzes}}, \bibinfo
  {author} {\bibfnamefont {J.~M.}\ \bibnamefont {Raimond}}, \bibinfo {author}
  {\bibfnamefont {S.}~\bibnamefont {Haroche}}, \bibinfo {author} {\bibfnamefont
  {I.}~\bibnamefont {Dotsenko}}, \ and\ \bibinfo {author} {\bibfnamefont
  {M.}~\bibnamefont {Brune}},\ }\bibinfo {title} {\emph {Adaptive {{Quantum
  Nondemolition Measurement}} of a {{Photon Number}}}},\ \href {\doibase
  10.1103/PhysRevLett.112.080401} {\bibfield  {journal} {\bibinfo  {journal}
  {Physical Review Letters}\ }\textbf {\bibinfo {volume} {112}},\ \bibinfo
  {pages} {080401} (\bibinfo {year} {2014})}\BibitemShut {NoStop}%
\bibitem [{\citenamefont {Royer}\ \emph {et~al.}(2018)\citenamefont {Royer},
  \citenamefont {Grimsmo}, \citenamefont {{Choquette-Poitevin}},\ and\
  \citenamefont {Blais}}]{royer_itinerant_2018}%
  \BibitemOpen
  \bibfield  {author} {\bibinfo {author} {\bibfnamefont {B.}~\bibnamefont
  {Royer}}, \bibinfo {author} {\bibfnamefont {A.~L.}\ \bibnamefont {Grimsmo}},
  \bibinfo {author} {\bibfnamefont {A.}~\bibnamefont {{Choquette-Poitevin}}}, \
  and\ \bibinfo {author} {\bibfnamefont {A.}~\bibnamefont {Blais}},\ }\bibinfo
  {title} {\emph {Itinerant {{Microwave Photon Detector}}}},\ \href {\doibase
  10.1103/PhysRevLett.120.203602} {\bibfield  {journal} {\bibinfo  {journal}
  {Physical Review Letters}\ }\textbf {\bibinfo {volume} {120}},\ \bibinfo
  {pages} {203602} (\bibinfo {year} {2018})}\BibitemShut {NoStop}%
\bibitem [{\citenamefont {Shahmoon}\ \emph {et~al.}(2011)\citenamefont
  {Shahmoon}, \citenamefont {Kurizki}, \citenamefont {Fleischhauer},\ and\
  \citenamefont {Petrosyan}}]{shahmoon_strongly_2011}%
  \BibitemOpen
  \bibfield  {author} {\bibinfo {author} {\bibfnamefont {E.}~\bibnamefont
  {Shahmoon}}, \bibinfo {author} {\bibfnamefont {G.}~\bibnamefont {Kurizki}},
  \bibinfo {author} {\bibfnamefont {M.}~\bibnamefont {Fleischhauer}}, \ and\
  \bibinfo {author} {\bibfnamefont {D.}~\bibnamefont {Petrosyan}},\ }\bibinfo
  {title} {\emph {Strongly Interacting Photons in Hollow-Core Waveguides}},\
  \href {\doibase 10.1103/PhysRevA.83.033806} {\bibfield  {journal} {\bibinfo
  {journal} {Physical Review A}\ }\textbf {\bibinfo {volume} {83}},\ \bibinfo
  {pages} {033806} (\bibinfo {year} {2011})}\BibitemShut {NoStop}%
\bibitem [{\citenamefont {Besse}\ \emph {et~al.}(2018)\citenamefont {Besse},
  \citenamefont {Gasparinetti}, \citenamefont {Collodo}, \citenamefont
  {Walter}, \citenamefont {Kurpiers}, \citenamefont {Pechal}, \citenamefont
  {Eichler},\ and\ \citenamefont {Wallraff}}]{besse_single-shot_2018}%
  \BibitemOpen
  \bibfield  {author} {\bibinfo {author} {\bibfnamefont {J.-C.}\ \bibnamefont
  {Besse}}, \bibinfo {author} {\bibfnamefont {S.}~\bibnamefont {Gasparinetti}},
  \bibinfo {author} {\bibfnamefont {M.~C.}\ \bibnamefont {Collodo}}, \bibinfo
  {author} {\bibfnamefont {T.}~\bibnamefont {Walter}}, \bibinfo {author}
  {\bibfnamefont {P.}~\bibnamefont {Kurpiers}}, \bibinfo {author}
  {\bibfnamefont {M.}~\bibnamefont {Pechal}}, \bibinfo {author} {\bibfnamefont
  {C.}~\bibnamefont {Eichler}}, \ and\ \bibinfo {author} {\bibfnamefont
  {A.}~\bibnamefont {Wallraff}},\ }\bibinfo {title} {\emph {Single-{{Shot
  Quantum Nondemolition Detection}} of {{Individual Itinerant Microwave
  Photons}}}},\ \href {\doibase 10.1103/PhysRevX.8.021003} {\bibfield
  {journal} {\bibinfo  {journal} {Physical Review X}\ }\textbf {\bibinfo
  {volume} {8}},\ \bibinfo {pages} {021003} (\bibinfo {year}
  {2018})}\BibitemShut {NoStop}%
\bibitem [{\citenamefont {Kono}\ \emph {et~al.}(2018)\citenamefont {Kono},
  \citenamefont {Koshino}, \citenamefont {Tabuchi}, \citenamefont {Noguchi},\
  and\ \citenamefont {Nakamura}}]{kono_quantum_2018}%
  \BibitemOpen
  \bibfield  {author} {\bibinfo {author} {\bibfnamefont {S.}~\bibnamefont
  {Kono}}, \bibinfo {author} {\bibfnamefont {K.}~\bibnamefont {Koshino}},
  \bibinfo {author} {\bibfnamefont {Y.}~\bibnamefont {Tabuchi}}, \bibinfo
  {author} {\bibfnamefont {A.}~\bibnamefont {Noguchi}}, \ and\ \bibinfo
  {author} {\bibfnamefont {Y.}~\bibnamefont {Nakamura}},\ }\bibinfo {title}
  {\emph {Quantum Non-Demolition Detection of an Itinerant Microwave Photon}},\
  \href {\doibase 10.1038/s41567-018-0066-3} {\bibfield  {journal} {\bibinfo
  {journal} {Nature Physics}\ }\textbf {\bibinfo {volume} {14}},\ \bibinfo
  {pages} {546} (\bibinfo {year} {2018})}\BibitemShut {NoStop}%
\bibitem [{\citenamefont {Gorshkov}\ \emph {et~al.}(2011)\citenamefont
  {Gorshkov}, \citenamefont {Otterbach}, \citenamefont {Fleischhauer},
  \citenamefont {Pohl},\ and\ \citenamefont
  {Lukin}}]{gorshkov_photon-photon_2011}%
  \BibitemOpen
  \bibfield  {author} {\bibinfo {author} {\bibfnamefont {A.~V.}\ \bibnamefont
  {Gorshkov}}, \bibinfo {author} {\bibfnamefont {J.}~\bibnamefont {Otterbach}},
  \bibinfo {author} {\bibfnamefont {M.}~\bibnamefont {Fleischhauer}}, \bibinfo
  {author} {\bibfnamefont {T.}~\bibnamefont {Pohl}}, \ and\ \bibinfo {author}
  {\bibfnamefont {M.~D.}\ \bibnamefont {Lukin}},\ }\bibinfo {title} {\emph
  {Photon-{{Photon Interactions}} via {{Rydberg Blockade}}}},\ \href {\doibase
  10.1103/PhysRevLett.107.133602} {\bibfield  {journal} {\bibinfo  {journal}
  {Physical Review Letters}\ }\textbf {\bibinfo {volume} {107}},\ \bibinfo
  {pages} {133602} (\bibinfo {year} {2011})}\BibitemShut {NoStop}%
\bibitem [{\citenamefont {Bienias}\ \emph {et~al.}(2014)\citenamefont
  {Bienias}, \citenamefont {Choi}, \citenamefont {Firstenberg}, \citenamefont
  {Maghrebi}, \citenamefont {Gullans}, \citenamefont {Lukin}, \citenamefont
  {Gorshkov},\ and\ \citenamefont {B{\"u}chler}}]{bienias_scattering_2014}%
  \BibitemOpen
  \bibfield  {author} {\bibinfo {author} {\bibfnamefont {P.}~\bibnamefont
  {Bienias}}, \bibinfo {author} {\bibfnamefont {S.}~\bibnamefont {Choi}},
  \bibinfo {author} {\bibfnamefont {O.}~\bibnamefont {Firstenberg}}, \bibinfo
  {author} {\bibfnamefont {M.~F.}\ \bibnamefont {Maghrebi}}, \bibinfo {author}
  {\bibfnamefont {M.}~\bibnamefont {Gullans}}, \bibinfo {author} {\bibfnamefont
  {M.~D.}\ \bibnamefont {Lukin}}, \bibinfo {author} {\bibfnamefont {A.~V.}\
  \bibnamefont {Gorshkov}}, \ and\ \bibinfo {author} {\bibfnamefont {H.~P.}\
  \bibnamefont {B{\"u}chler}},\ }\bibinfo {title} {\emph {Scattering Resonances
  and Bound States for Strongly Interacting {{Rydberg}} Polaritons}},\ \href
  {\doibase 10.1103/PhysRevA.90.053804} {\bibfield  {journal} {\bibinfo
  {journal} {Physical Review A}\ }\textbf {\bibinfo {volume} {90}},\ \bibinfo
  {pages} {053804} (\bibinfo {year} {2014})}\BibitemShut {NoStop}%
\bibitem [{\citenamefont {Zeuthen}\ \emph {et~al.}(2017)\citenamefont
  {Zeuthen}, \citenamefont {Gullans}, \citenamefont {Maghrebi},\ and\
  \citenamefont {Gorshkov}}]{zeuthen_correlated_2017}%
  \BibitemOpen
  \bibfield  {author} {\bibinfo {author} {\bibfnamefont {E.}~\bibnamefont
  {Zeuthen}}, \bibinfo {author} {\bibfnamefont {M.~J.}\ \bibnamefont
  {Gullans}}, \bibinfo {author} {\bibfnamefont {M.~F.}\ \bibnamefont
  {Maghrebi}}, \ and\ \bibinfo {author} {\bibfnamefont {A.~V.}\ \bibnamefont
  {Gorshkov}},\ }\bibinfo {title} {\emph {Correlated {{Photon Dynamics}} in
  {{Dissipative Rydberg Media}}}},\ \href {\doibase
  10.1103/PhysRevLett.119.043602} {\bibfield  {journal} {\bibinfo  {journal}
  {Physical Review Letters}\ }\textbf {\bibinfo {volume} {119}},\ \bibinfo
  {pages} {043602} (\bibinfo {year} {2017})}\BibitemShut {NoStop}%
\bibitem [{\citenamefont {Bienias}\ \emph {et~al.}(2020)\citenamefont
  {Bienias}, \citenamefont {Gullans}, \citenamefont {Kalinowski}, \citenamefont
  {Craddock}, \citenamefont {{Ornelas-Huerta}}, \citenamefont {Rolston},
  \citenamefont {Porto},\ and\ \citenamefont {Gorshkov}}]{bienias_exotic_2020}%
  \BibitemOpen
  \bibfield  {author} {\bibinfo {author} {\bibfnamefont {P.}~\bibnamefont
  {Bienias}}, \bibinfo {author} {\bibfnamefont {M.~J.}\ \bibnamefont
  {Gullans}}, \bibinfo {author} {\bibfnamefont {M.}~\bibnamefont {Kalinowski}},
  \bibinfo {author} {\bibfnamefont {A.~N.}\ \bibnamefont {Craddock}}, \bibinfo
  {author} {\bibfnamefont {D.~P.}\ \bibnamefont {{Ornelas-Huerta}}}, \bibinfo
  {author} {\bibfnamefont {S.~L.}\ \bibnamefont {Rolston}}, \bibinfo {author}
  {\bibfnamefont {J.~V.}\ \bibnamefont {Porto}}, \ and\ \bibinfo {author}
  {\bibfnamefont {A.~V.}\ \bibnamefont {Gorshkov}},\ }\bibinfo {title} {\emph
  {Exotic {{Photonic Molecules}} via {{Lennard-Jones-like Potentials}}}},\
  \href {\doibase 10.1103/PhysRevLett.125.093601} {\bibfield  {journal}
  {\bibinfo  {journal} {Physical Review Letters}\ }\textbf {\bibinfo {volume}
  {125}},\ \bibinfo {pages} {093601} (\bibinfo {year} {2020})}\BibitemShut
  {NoStop}%
\bibitem [{\citenamefont {{Ornelas-Huerta}}\ \emph {et~al.}(2021)\citenamefont
  {{Ornelas-Huerta}}, \citenamefont {Bienias}, \citenamefont {Craddock},
  \citenamefont {Gullans}, \citenamefont {Hachtel}, \citenamefont {Kalinowski},
  \citenamefont {Lyon}, \citenamefont {Gorshkov}, \citenamefont {Rolston},\
  and\ \citenamefont {Porto}}]{ornelas-huerta_tunable_2021}%
  \BibitemOpen
  \bibfield  {author} {\bibinfo {author} {\bibfnamefont {D.~P.}\ \bibnamefont
  {{Ornelas-Huerta}}}, \bibinfo {author} {\bibfnamefont {P.}~\bibnamefont
  {Bienias}}, \bibinfo {author} {\bibfnamefont {A.~N.}\ \bibnamefont
  {Craddock}}, \bibinfo {author} {\bibfnamefont {M.~J.}\ \bibnamefont
  {Gullans}}, \bibinfo {author} {\bibfnamefont {A.~J.}\ \bibnamefont
  {Hachtel}}, \bibinfo {author} {\bibfnamefont {M.}~\bibnamefont {Kalinowski}},
  \bibinfo {author} {\bibfnamefont {M.~E.}\ \bibnamefont {Lyon}}, \bibinfo
  {author} {\bibfnamefont {A.~V.}\ \bibnamefont {Gorshkov}}, \bibinfo {author}
  {\bibfnamefont {S.~L.}\ \bibnamefont {Rolston}}, \ and\ \bibinfo {author}
  {\bibfnamefont {J.~V.}\ \bibnamefont {Porto}},\ }\bibinfo {title} {\emph
  {Tunable {{Three-Body Loss}} in a {{Nonlinear Rydberg Medium}}}},\ \href
  {\doibase 10.1103/PhysRevLett.126.173401} {\bibfield  {journal} {\bibinfo
  {journal} {Physical Review Letters}\ }\textbf {\bibinfo {volume} {126}},\
  \bibinfo {pages} {173401} (\bibinfo {year} {2021})}\BibitemShut {NoStop}%
\bibitem [{\citenamefont {Maghrebi}\ \emph {et~al.}(2015)\citenamefont
  {Maghrebi}, \citenamefont {Yao}, \citenamefont {Hafezi}, \citenamefont
  {Pohl}, \citenamefont {Firstenberg},\ and\ \citenamefont
  {Gorshkov}}]{maghrebi_fractional_2015}%
  \BibitemOpen
  \bibfield  {author} {\bibinfo {author} {\bibfnamefont {M.~F.}\ \bibnamefont
  {Maghrebi}}, \bibinfo {author} {\bibfnamefont {N.~Y.}\ \bibnamefont {Yao}},
  \bibinfo {author} {\bibfnamefont {M.}~\bibnamefont {Hafezi}}, \bibinfo
  {author} {\bibfnamefont {T.}~\bibnamefont {Pohl}}, \bibinfo {author}
  {\bibfnamefont {O.}~\bibnamefont {Firstenberg}}, \ and\ \bibinfo {author}
  {\bibfnamefont {A.~V.}\ \bibnamefont {Gorshkov}},\ }\bibinfo {title} {\emph
  {Fractional Quantum {{Hall}} States of {{Rydberg}} Polaritons}},\ \href
  {\doibase 10.1103/PhysRevA.91.033838} {\bibfield  {journal} {\bibinfo
  {journal} {Physical Review A}\ }\textbf {\bibinfo {volume} {91}},\ \bibinfo
  {pages} {033838} (\bibinfo {year} {2015})}\BibitemShut {NoStop}%
\bibitem [{\citenamefont {Omran}\ \emph {et~al.}(2019)\citenamefont {Omran},
  \citenamefont {Levine}, \citenamefont {Keesling}, \citenamefont {Semeghini},
  \citenamefont {Wang}, \citenamefont {Ebadi}, \citenamefont {Bernien},
  \citenamefont {Zibrov}, \citenamefont {Pichler}, \citenamefont {Choi},
  \citenamefont {Cui}, \citenamefont {Rossignolo}, \citenamefont {Rembold},
  \citenamefont {Montangero}, \citenamefont {Calarco}, \citenamefont {Endres},
  \citenamefont {Greiner}, \citenamefont {Vuleti{\'c}},\ and\ \citenamefont
  {Lukin}}]{omran_generation_2019}%
  \BibitemOpen
  \bibfield  {author} {\bibinfo {author} {\bibfnamefont {A.}~\bibnamefont
  {Omran}}, \bibinfo {author} {\bibfnamefont {H.}~\bibnamefont {Levine}},
  \bibinfo {author} {\bibfnamefont {A.}~\bibnamefont {Keesling}}, \bibinfo
  {author} {\bibfnamefont {G.}~\bibnamefont {Semeghini}}, \bibinfo {author}
  {\bibfnamefont {T.~T.}\ \bibnamefont {Wang}}, \bibinfo {author}
  {\bibfnamefont {S.}~\bibnamefont {Ebadi}}, \bibinfo {author} {\bibfnamefont
  {H.}~\bibnamefont {Bernien}}, \bibinfo {author} {\bibfnamefont {A.~S.}\
  \bibnamefont {Zibrov}}, \bibinfo {author} {\bibfnamefont {H.}~\bibnamefont
  {Pichler}}, \bibinfo {author} {\bibfnamefont {S.}~\bibnamefont {Choi}},
  \bibinfo {author} {\bibfnamefont {J.}~\bibnamefont {Cui}}, \bibinfo {author}
  {\bibfnamefont {M.}~\bibnamefont {Rossignolo}}, \bibinfo {author}
  {\bibfnamefont {P.}~\bibnamefont {Rembold}}, \bibinfo {author} {\bibfnamefont
  {S.}~\bibnamefont {Montangero}}, \bibinfo {author} {\bibfnamefont
  {T.}~\bibnamefont {Calarco}}, \bibinfo {author} {\bibfnamefont
  {M.}~\bibnamefont {Endres}}, \bibinfo {author} {\bibfnamefont
  {M.}~\bibnamefont {Greiner}}, \bibinfo {author} {\bibfnamefont
  {V.}~\bibnamefont {Vuleti{\'c}}}, \ and\ \bibinfo {author} {\bibfnamefont
  {M.~D.}\ \bibnamefont {Lukin}},\ }\bibinfo {title} {\emph {Generation and
  Manipulation of {{Schr\"odinger}} Cat States in {{Rydberg}} Atom Arrays}},\
  \href {\doibase 10.1126/science.aax9743} {\bibfield  {journal} {\bibinfo
  {journal} {Science}\ }\textbf {\bibinfo {volume} {365}},\ \bibinfo {pages}
  {570} (\bibinfo {year} {2019})}\BibitemShut {NoStop}%
\bibitem [{\citenamefont {Celi}\ \emph {et~al.}(2020)\citenamefont {Celi},
  \citenamefont {Vermersch}, \citenamefont {Viyuela}, \citenamefont {Pichler},
  \citenamefont {Lukin},\ and\ \citenamefont {Zoller}}]{celi_emerging_2020}%
  \BibitemOpen
  \bibfield  {author} {\bibinfo {author} {\bibfnamefont {A.}~\bibnamefont
  {Celi}}, \bibinfo {author} {\bibfnamefont {B.}~\bibnamefont {Vermersch}},
  \bibinfo {author} {\bibfnamefont {O.}~\bibnamefont {Viyuela}}, \bibinfo
  {author} {\bibfnamefont {H.}~\bibnamefont {Pichler}}, \bibinfo {author}
  {\bibfnamefont {M.~D.}\ \bibnamefont {Lukin}}, \ and\ \bibinfo {author}
  {\bibfnamefont {P.}~\bibnamefont {Zoller}},\ }\bibinfo {title} {\emph
  {Emerging {{Two-Dimensional Gauge Theories}} in {{Rydberg Configurable
  Arrays}}}},\ \href {\doibase 10.1103/PhysRevX.10.021057} {\bibfield
  {journal} {\bibinfo  {journal} {Physical Review X}\ }\textbf {\bibinfo
  {volume} {10}},\ \bibinfo {pages} {021057} (\bibinfo {year}
  {2020})}\BibitemShut {NoStop}%
\bibitem [{\citenamefont {Samajdar}\ \emph {et~al.}(2020)\citenamefont
  {Samajdar}, \citenamefont {Ho}, \citenamefont {Pichler}, \citenamefont
  {Lukin},\ and\ \citenamefont {Sachdev}}]{samajdar_complex_2020}%
  \BibitemOpen
  \bibfield  {author} {\bibinfo {author} {\bibfnamefont {R.}~\bibnamefont
  {Samajdar}}, \bibinfo {author} {\bibfnamefont {W.~W.}\ \bibnamefont {Ho}},
  \bibinfo {author} {\bibfnamefont {H.}~\bibnamefont {Pichler}}, \bibinfo
  {author} {\bibfnamefont {M.~D.}\ \bibnamefont {Lukin}}, \ and\ \bibinfo
  {author} {\bibfnamefont {S.}~\bibnamefont {Sachdev}},\ }\bibinfo {title}
  {\emph {Complex {{Density Wave Orders}} and {{Quantum Phase Transitions}} in
  a {{Model}} of {{Square-Lattice Rydberg Atom Arrays}}}},\ \href {\doibase
  10.1103/PhysRevLett.124.103601} {\bibfield  {journal} {\bibinfo  {journal}
  {Physical Review Letters}\ }\textbf {\bibinfo {volume} {124}},\ \bibinfo
  {pages} {103601} (\bibinfo {year} {2020})}\BibitemShut {NoStop}%
\bibitem [{\citenamefont {Semeghini}\ \emph {et~al.}(2021)\citenamefont
  {Semeghini}, \citenamefont {Levine}, \citenamefont {Keesling}, \citenamefont
  {Ebadi}, \citenamefont {Wang}, \citenamefont {Bluvstein}, \citenamefont
  {Verresen}, \citenamefont {Pichler}, \citenamefont {Kalinowski},
  \citenamefont {Samajdar}, \citenamefont {Omran}, \citenamefont {Sachdev},
  \citenamefont {Vishwanath}, \citenamefont {Greiner}, \citenamefont
  {Vuleti{\'c}},\ and\ \citenamefont {Lukin}}]{semeghini_probing_2021}%
  \BibitemOpen
  \bibfield  {author} {\bibinfo {author} {\bibfnamefont {G.}~\bibnamefont
  {Semeghini}}, \bibinfo {author} {\bibfnamefont {H.}~\bibnamefont {Levine}},
  \bibinfo {author} {\bibfnamefont {A.}~\bibnamefont {Keesling}}, \bibinfo
  {author} {\bibfnamefont {S.}~\bibnamefont {Ebadi}}, \bibinfo {author}
  {\bibfnamefont {T.~T.}\ \bibnamefont {Wang}}, \bibinfo {author}
  {\bibfnamefont {D.}~\bibnamefont {Bluvstein}}, \bibinfo {author}
  {\bibfnamefont {R.}~\bibnamefont {Verresen}}, \bibinfo {author}
  {\bibfnamefont {H.}~\bibnamefont {Pichler}}, \bibinfo {author} {\bibfnamefont
  {M.}~\bibnamefont {Kalinowski}}, \bibinfo {author} {\bibfnamefont
  {R.}~\bibnamefont {Samajdar}}, \bibinfo {author} {\bibfnamefont
  {A.}~\bibnamefont {Omran}}, \bibinfo {author} {\bibfnamefont
  {S.}~\bibnamefont {Sachdev}}, \bibinfo {author} {\bibfnamefont
  {A.}~\bibnamefont {Vishwanath}}, \bibinfo {author} {\bibfnamefont
  {M.}~\bibnamefont {Greiner}}, \bibinfo {author} {\bibfnamefont
  {V.}~\bibnamefont {Vuleti{\'c}}}, \ and\ \bibinfo {author} {\bibfnamefont
  {M.~D.}\ \bibnamefont {Lukin}},\ }\bibinfo {title} {\emph {Probing
  Topological Spin Liquids on a Programmable Quantum Simulator}},\ \href
  {\doibase 10.1126/science.abi8794} {\bibfield  {journal} {\bibinfo  {journal}
  {Science}\ }\textbf {\bibinfo {volume} {374}},\ \bibinfo {pages} {1242}
  (\bibinfo {year} {2021})}\BibitemShut {NoStop}%
\bibitem [{\citenamefont {Kalinowski}\ \emph {et~al.}(2022)\citenamefont
  {Kalinowski}, \citenamefont {Samajdar}, \citenamefont {Melko}, \citenamefont
  {Lukin}, \citenamefont {Sachdev},\ and\ \citenamefont
  {Choi}}]{kalinowski_bulk_2022}%
  \BibitemOpen
  \bibfield  {author} {\bibinfo {author} {\bibfnamefont {M.}~\bibnamefont
  {Kalinowski}}, \bibinfo {author} {\bibfnamefont {R.}~\bibnamefont
  {Samajdar}}, \bibinfo {author} {\bibfnamefont {R.~G.}\ \bibnamefont {Melko}},
  \bibinfo {author} {\bibfnamefont {M.~D.}\ \bibnamefont {Lukin}}, \bibinfo
  {author} {\bibfnamefont {S.}~\bibnamefont {Sachdev}}, \ and\ \bibinfo
  {author} {\bibfnamefont {S.}~\bibnamefont {Choi}},\ }\bibinfo {title} {\emph
  {Bulk and Boundary Quantum Phase Transitions in a Square {{Rydberg}} Atom
  Array}},\ \href {\doibase 10.1103/PhysRevB.105.174417} {\bibfield  {journal}
  {\bibinfo  {journal} {Physical Review B}\ }\textbf {\bibinfo {volume}
  {105}},\ \bibinfo {pages} {174417} (\bibinfo {year} {2022})}\BibitemShut
  {NoStop}%
\bibitem [{\citenamefont {Bluvstein}\ \emph {et~al.}(2022)\citenamefont
  {Bluvstein}, \citenamefont {Levine}, \citenamefont {Semeghini}, \citenamefont
  {Wang}, \citenamefont {Ebadi}, \citenamefont {Kalinowski}, \citenamefont
  {Keesling}, \citenamefont {Maskara}, \citenamefont {Pichler}, \citenamefont
  {Greiner}, \citenamefont {Vuleti{\'c}},\ and\ \citenamefont
  {Lukin}}]{bluvstein_quantum_2022}%
  \BibitemOpen
  \bibfield  {author} {\bibinfo {author} {\bibfnamefont {D.}~\bibnamefont
  {Bluvstein}}, \bibinfo {author} {\bibfnamefont {H.}~\bibnamefont {Levine}},
  \bibinfo {author} {\bibfnamefont {G.}~\bibnamefont {Semeghini}}, \bibinfo
  {author} {\bibfnamefont {T.~T.}\ \bibnamefont {Wang}}, \bibinfo {author}
  {\bibfnamefont {S.}~\bibnamefont {Ebadi}}, \bibinfo {author} {\bibfnamefont
  {M.}~\bibnamefont {Kalinowski}}, \bibinfo {author} {\bibfnamefont
  {A.}~\bibnamefont {Keesling}}, \bibinfo {author} {\bibfnamefont
  {N.}~\bibnamefont {Maskara}}, \bibinfo {author} {\bibfnamefont
  {H.}~\bibnamefont {Pichler}}, \bibinfo {author} {\bibfnamefont
  {M.}~\bibnamefont {Greiner}}, \bibinfo {author} {\bibfnamefont
  {V.}~\bibnamefont {Vuleti{\'c}}}, \ and\ \bibinfo {author} {\bibfnamefont
  {M.~D.}\ \bibnamefont {Lukin}},\ }\bibinfo {title} {\emph {A Quantum
  Processor Based on Coherent Transport of Entangled Atom Arrays}},\ \href
  {\doibase 10.1038/s41586-022-04592-6} {\bibfield  {journal} {\bibinfo
  {journal} {Nature}\ }\textbf {\bibinfo {volume} {604}},\ \bibinfo {pages}
  {451} (\bibinfo {year} {2022})}\BibitemShut {NoStop}%
\bibitem [{\citenamefont {Sun}\ and\ \citenamefont
  {Chen}(2018)}]{sun_analysis_2018}%
  \BibitemOpen
  \bibfield  {author} {\bibinfo {author} {\bibfnamefont {Y.}~\bibnamefont
  {Sun}}\ and\ \bibinfo {author} {\bibfnamefont {P.-X.}\ \bibnamefont {Chen}},\
  }\bibinfo {title} {\emph {Analysis of Atom\&\#x2013;Photon Quantum Interface
  with Intracavity {{Rydberg-blocked}} Atomic Ensemble via Two-Photon
  Transition}},\ \href {\doibase 10.1364/OPTICA.5.001492} {\bibfield  {journal}
  {\bibinfo  {journal} {Optica}\ }\textbf {\bibinfo {volume} {5}},\ \bibinfo
  {pages} {1492} (\bibinfo {year} {2018})}\BibitemShut {NoStop}%
\bibitem [{\citenamefont {Ripka}\ \emph {et~al.}(2018)\citenamefont {Ripka},
  \citenamefont {K{\"u}bler}, \citenamefont {L{\"o}w},\ and\ \citenamefont
  {Pfau}}]{ripka_room-temperature_2018}%
  \BibitemOpen
  \bibfield  {author} {\bibinfo {author} {\bibfnamefont {F.}~\bibnamefont
  {Ripka}}, \bibinfo {author} {\bibfnamefont {H.}~\bibnamefont {K{\"u}bler}},
  \bibinfo {author} {\bibfnamefont {R.}~\bibnamefont {L{\"o}w}}, \ and\
  \bibinfo {author} {\bibfnamefont {T.}~\bibnamefont {Pfau}},\ }\bibinfo
  {title} {\emph {A Room-Temperature Single-Photon Source Based on Strongly
  Interacting {{Rydberg}} Atoms}},\ \href {\doibase 10.1126/science.aau1949}
  {\bibfield  {journal} {\bibinfo  {journal} {Science}\ }\textbf {\bibinfo
  {volume} {362}},\ \bibinfo {pages} {446} (\bibinfo {year}
  {2018})}\BibitemShut {NoStop}%
\bibitem [{\citenamefont {Tiarks}\ \emph {et~al.}(2019)\citenamefont {Tiarks},
  \citenamefont {{Schmidt-Eberle}}, \citenamefont {Stolz}, \citenamefont
  {Rempe},\ and\ \citenamefont {D{\"u}rr}}]{tiarks_photonphoton_2019}%
  \BibitemOpen
  \bibfield  {author} {\bibinfo {author} {\bibfnamefont {D.}~\bibnamefont
  {Tiarks}}, \bibinfo {author} {\bibfnamefont {S.}~\bibnamefont
  {{Schmidt-Eberle}}}, \bibinfo {author} {\bibfnamefont {T.}~\bibnamefont
  {Stolz}}, \bibinfo {author} {\bibfnamefont {G.}~\bibnamefont {Rempe}}, \ and\
  \bibinfo {author} {\bibfnamefont {S.}~\bibnamefont {D{\"u}rr}},\ }\bibinfo
  {title} {\emph {A Photon\textendash Photon Quantum Gate Based on {{Rydberg}}
  Interactions}},\ \href {\doibase 10.1038/s41567-018-0313-7} {\bibfield
  {journal} {\bibinfo  {journal} {Nature Physics}\ }\textbf {\bibinfo {volume}
  {15}},\ \bibinfo {pages} {124} (\bibinfo {year} {2019})}\BibitemShut
  {NoStop}%
\bibitem [{\citenamefont {{Ornelas-Huerta}}\ \emph {et~al.}(2020)\citenamefont
  {{Ornelas-Huerta}}, \citenamefont {Craddock}, \citenamefont {Goldschmidt},
  \citenamefont {Goldschmidt}, \citenamefont {Hachtel}, \citenamefont {Wang},
  \citenamefont {Bienias}, \citenamefont {Bienias}, \citenamefont {Gorshkov},
  \citenamefont {Gorshkov}, \citenamefont {Rolston},\ and\ \citenamefont
  {Porto}}]{ornelas-huerta_-demand_2020}%
  \BibitemOpen
  \bibfield  {author} {\bibinfo {author} {\bibfnamefont {D.~P.}\ \bibnamefont
  {{Ornelas-Huerta}}}, \bibinfo {author} {\bibfnamefont {A.~N.}\ \bibnamefont
  {Craddock}}, \bibinfo {author} {\bibfnamefont {E.~A.}\ \bibnamefont
  {Goldschmidt}}, \bibinfo {author} {\bibfnamefont {E.~A.}\ \bibnamefont
  {Goldschmidt}}, \bibinfo {author} {\bibfnamefont {A.~J.}\ \bibnamefont
  {Hachtel}}, \bibinfo {author} {\bibfnamefont {Y.}~\bibnamefont {Wang}},
  \bibinfo {author} {\bibfnamefont {P.}~\bibnamefont {Bienias}}, \bibinfo
  {author} {\bibfnamefont {P.}~\bibnamefont {Bienias}}, \bibinfo {author}
  {\bibfnamefont {A.~V.}\ \bibnamefont {Gorshkov}}, \bibinfo {author}
  {\bibfnamefont {A.~V.}\ \bibnamefont {Gorshkov}}, \bibinfo {author}
  {\bibfnamefont {S.~L.}\ \bibnamefont {Rolston}}, \ and\ \bibinfo {author}
  {\bibfnamefont {J.~V.}\ \bibnamefont {Porto}},\ }\bibinfo {title} {\emph
  {On-Demand Indistinguishable Single Photons from an Efficient and Pure Source
  Based on a {{Rydberg}} Ensemble}},\ \href {\doibase 10.1364/OPTICA.391485}
  {\bibfield  {journal} {\bibinfo  {journal} {Optica}\ }\textbf {\bibinfo
  {volume} {7}},\ \bibinfo {pages} {813} (\bibinfo {year} {2020})}\BibitemShut
  {NoStop}%
\bibitem [{\citenamefont {Bekenstein}\ \emph {et~al.}(2020)\citenamefont
  {Bekenstein}, \citenamefont {Pikovski}, \citenamefont {Pichler},
  \citenamefont {Shahmoon}, \citenamefont {Yelin},\ and\ \citenamefont
  {Lukin}}]{bekenstein_quantum_2020}%
  \BibitemOpen
  \bibfield  {author} {\bibinfo {author} {\bibfnamefont {R.}~\bibnamefont
  {Bekenstein}}, \bibinfo {author} {\bibfnamefont {I.}~\bibnamefont
  {Pikovski}}, \bibinfo {author} {\bibfnamefont {H.}~\bibnamefont {Pichler}},
  \bibinfo {author} {\bibfnamefont {E.}~\bibnamefont {Shahmoon}}, \bibinfo
  {author} {\bibfnamefont {S.~F.}\ \bibnamefont {Yelin}}, \ and\ \bibinfo
  {author} {\bibfnamefont {M.~D.}\ \bibnamefont {Lukin}},\ }\bibinfo {title}
  {\emph {Quantum Metasurfaces with Atom Arrays}},\ \href {\doibase
  10.1038/s41567-020-0845-5} {\bibfield  {journal} {\bibinfo  {journal} {Nature
  Physics}\ }\textbf {\bibinfo {volume} {16}},\ \bibinfo {pages} {676}
  (\bibinfo {year} {2020})}\BibitemShut {NoStop}%
\bibitem [{\citenamefont {Zhang}\ \emph {et~al.}(2021)\citenamefont {Zhang},
  \citenamefont {Walther}, \citenamefont {M{\o}lmer},\ and\ \citenamefont
  {Pohl}}]{zhang_photon-photon_2021}%
  \BibitemOpen
  \bibfield  {author} {\bibinfo {author} {\bibfnamefont {L.}~\bibnamefont
  {Zhang}}, \bibinfo {author} {\bibfnamefont {V.}~\bibnamefont {Walther}},
  \bibinfo {author} {\bibfnamefont {K.}~\bibnamefont {M{\o}lmer}}, \ and\
  \bibinfo {author} {\bibfnamefont {T.}~\bibnamefont {Pohl}},\ }\bibinfo
  {title} {\emph {Photon-Photon Interactions in {{Rydberg-atom}} Arrays}},\
  \href@noop {} {\bibfield  {journal} {\bibinfo  {journal} {arXiv:2101.11375
  [quant-ph]}\ } (\bibinfo {year} {2021})},\ \Eprint
  {http://arxiv.org/abs/2101.11375} {arXiv:2101.11375 [quant-ph]}\BibitemShut
  {NoStop}%
\bibitem [{\citenamefont {{Moreno-Cardoner}}\ \emph {et~al.}(2021)\citenamefont
  {{Moreno-Cardoner}}, \citenamefont {Goncalves},\ and\ \citenamefont
  {Chang}}]{moreno-cardoner_quantum_2021}%
  \BibitemOpen
  \bibfield  {author} {\bibinfo {author} {\bibfnamefont {M.}~\bibnamefont
  {{Moreno-Cardoner}}}, \bibinfo {author} {\bibfnamefont {D.}~\bibnamefont
  {Goncalves}}, \ and\ \bibinfo {author} {\bibfnamefont {D.~E.}\ \bibnamefont
  {Chang}},\ }\bibinfo {title} {\emph {Quantum Nonlinear Optics Based on
  Two-Dimensional {{Rydberg}} Atom Arrays}},\ \href@noop {} {\bibfield
  {journal} {\bibinfo  {journal} {arXiv:2101.01936 [quant-ph]}\ } (\bibinfo
  {year} {2021})},\ \Eprint {http://arxiv.org/abs/2101.01936} {arXiv:2101.01936
  [quant-ph]}\BibitemShut {NoStop}%
\bibitem [{\citenamefont {Pedersen}\ \emph {et~al.}(2022)\citenamefont
  {Pedersen}, \citenamefont {Zhang},\ and\ \citenamefont
  {Pohl}}]{pedersen_quantum_2022-1}%
  \BibitemOpen
  \bibfield  {author} {\bibinfo {author} {\bibfnamefont {S.~P.}\ \bibnamefont
  {Pedersen}}, \bibinfo {author} {\bibfnamefont {L.}~\bibnamefont {Zhang}}, \
  and\ \bibinfo {author} {\bibfnamefont {T.}~\bibnamefont {Pohl}},\ }\href
  {\doibase 10.48550/arXiv.2201.06544} {\bibinfo {title} {\emph {Quantum
  Nonlinear Optics in Atomic Dual Arrays}},\ } (\bibinfo {year} {2022}),\
  \Eprint {http://arxiv.org/abs/2201.06544} {arXiv:2201.06544
  [quant-ph]}\BibitemShut {NoStop}%
\bibitem [{\citenamefont {Srakaew}\ \emph {et~al.}(2022)\citenamefont
  {Srakaew}, \citenamefont {Weckesser}, \citenamefont {Hollerith},
  \citenamefont {Wei}, \citenamefont {Adler}, \citenamefont {Bloch},\ and\
  \citenamefont {Zeiher}}]{srakaew_subwavelength_2022}%
  \BibitemOpen
  \bibfield  {author} {\bibinfo {author} {\bibfnamefont {K.}~\bibnamefont
  {Srakaew}}, \bibinfo {author} {\bibfnamefont {P.}~\bibnamefont {Weckesser}},
  \bibinfo {author} {\bibfnamefont {S.}~\bibnamefont {Hollerith}}, \bibinfo
  {author} {\bibfnamefont {D.}~\bibnamefont {Wei}}, \bibinfo {author}
  {\bibfnamefont {D.}~\bibnamefont {Adler}}, \bibinfo {author} {\bibfnamefont
  {I.}~\bibnamefont {Bloch}}, \ and\ \bibinfo {author} {\bibfnamefont
  {J.}~\bibnamefont {Zeiher}},\ }\href {\doibase 10.48550/arXiv.2207.09383}
  {\bibinfo {title} {\emph {A Subwavelength Atomic Array Switched by a Single
  {{Rydberg}} Atom}},\ } (\bibinfo {year} {2022}),\ \Eprint
  {http://arxiv.org/abs/2207.09383} {arXiv:2207.09383 [cond-mat,
  physics:physics, physics:quant-ph]}\BibitemShut {NoStop}%
\bibitem [{\citenamefont {Bettles}\ \emph {et~al.}(2016)\citenamefont
  {Bettles}, \citenamefont {Gardiner},\ and\ \citenamefont
  {Adams}}]{bettles_enhanced_2016}%
  \BibitemOpen
  \bibfield  {author} {\bibinfo {author} {\bibfnamefont {R.~J.}\ \bibnamefont
  {Bettles}}, \bibinfo {author} {\bibfnamefont {S.~A.}\ \bibnamefont
  {Gardiner}}, \ and\ \bibinfo {author} {\bibfnamefont {C.~S.}\ \bibnamefont
  {Adams}},\ }\bibinfo {title} {\emph {Enhanced {{Optical Cross Section}} via
  {{Collective Coupling}} of {{Atomic Dipoles}} in a {{2D Array}}}},\ \href
  {\doibase 10.1103/PhysRevLett.116.103602} {\bibfield  {journal} {\bibinfo
  {journal} {Physical Review Letters}\ }\textbf {\bibinfo {volume} {116}},\
  \bibinfo {pages} {103602} (\bibinfo {year} {2016})}\BibitemShut {NoStop}%
\bibitem [{\citenamefont {Shahmoon}\ \emph {et~al.}(2017)\citenamefont
  {Shahmoon}, \citenamefont {Wild}, \citenamefont {Lukin},\ and\ \citenamefont
  {Yelin}}]{shahmoon_cooperative_2017}%
  \BibitemOpen
  \bibfield  {author} {\bibinfo {author} {\bibfnamefont {E.}~\bibnamefont
  {Shahmoon}}, \bibinfo {author} {\bibfnamefont {D.~S.}\ \bibnamefont {Wild}},
  \bibinfo {author} {\bibfnamefont {M.~D.}\ \bibnamefont {Lukin}}, \ and\
  \bibinfo {author} {\bibfnamefont {S.~F.}\ \bibnamefont {Yelin}},\ }\bibinfo
  {title} {\emph {Cooperative {{Resonances}} in {{Light Scattering}} from
  {{Two-Dimensional Atomic Arrays}}}},\ \href {\doibase
  10.1103/PhysRevLett.118.113601} {\bibfield  {journal} {\bibinfo  {journal}
  {Physical Review Letters}\ }\textbf {\bibinfo {volume} {118}},\ \bibinfo
  {pages} {113601} (\bibinfo {year} {2017})}\BibitemShut {NoStop}%
\bibitem [{\citenamefont {Rui}\ \emph {et~al.}(2020)\citenamefont {Rui},
  \citenamefont {Wei}, \citenamefont {{Rubio-Abadal}}, \citenamefont
  {Hollerith}, \citenamefont {Zeiher}, \citenamefont {{Stamper-Kurn}},
  \citenamefont {Gross},\ and\ \citenamefont {Bloch}}]{rui_subradiant_2020}%
  \BibitemOpen
  \bibfield  {author} {\bibinfo {author} {\bibfnamefont {J.}~\bibnamefont
  {Rui}}, \bibinfo {author} {\bibfnamefont {D.}~\bibnamefont {Wei}}, \bibinfo
  {author} {\bibfnamefont {A.}~\bibnamefont {{Rubio-Abadal}}}, \bibinfo
  {author} {\bibfnamefont {S.}~\bibnamefont {Hollerith}}, \bibinfo {author}
  {\bibfnamefont {J.}~\bibnamefont {Zeiher}}, \bibinfo {author} {\bibfnamefont
  {D.~M.}\ \bibnamefont {{Stamper-Kurn}}}, \bibinfo {author} {\bibfnamefont
  {C.}~\bibnamefont {Gross}}, \ and\ \bibinfo {author} {\bibfnamefont
  {I.}~\bibnamefont {Bloch}},\ }\bibinfo {title} {\emph {A Subradiant Optical
  Mirror Formed by a Single Structured Atomic Layer}},\ \href {\doibase
  10.1038/s41586-020-2463-x} {\bibfield  {journal} {\bibinfo  {journal}
  {Nature}\ }\textbf {\bibinfo {volume} {583}},\ \bibinfo {pages} {369}
  (\bibinfo {year} {2020})}\BibitemShut {NoStop}%
\bibitem [{\citenamefont {Manzoni}\ \emph {et~al.}(2018)\citenamefont
  {Manzoni}, \citenamefont {{Moreno-Cardoner}}, \citenamefont
  {{Asenjo-Garcia}}, \citenamefont {Porto}, \citenamefont {Gorshkov},\ and\
  \citenamefont {Chang}}]{manzoni_optimization_2018}%
  \BibitemOpen
  \bibfield  {author} {\bibinfo {author} {\bibfnamefont {M.~T.}\ \bibnamefont
  {Manzoni}}, \bibinfo {author} {\bibfnamefont {M.}~\bibnamefont
  {{Moreno-Cardoner}}}, \bibinfo {author} {\bibfnamefont {A.}~\bibnamefont
  {{Asenjo-Garcia}}}, \bibinfo {author} {\bibfnamefont {J.~V.}\ \bibnamefont
  {Porto}}, \bibinfo {author} {\bibfnamefont {A.~V.}\ \bibnamefont {Gorshkov}},
  \ and\ \bibinfo {author} {\bibfnamefont {D.~E.}\ \bibnamefont {Chang}},\
  }\bibinfo {title} {\emph {Optimization of Photon Storage Fidelity in Ordered
  Atomic Arrays}},\ \href {\doibase 10.1088/1367-2630/aadb74} {\bibfield
  {journal} {\bibinfo  {journal} {New Journal of Physics}\ }\textbf {\bibinfo
  {volume} {20}},\ \bibinfo {pages} {083048} (\bibinfo {year} {2018})},\
  \Eprint {http://arxiv.org/abs/1710.06312} {arXiv:1710.06312}\BibitemShut
  {NoStop}%
\bibitem [{\citenamefont {Wei}\ \emph {et~al.}(2021)\citenamefont {Wei},
  \citenamefont {Malz}, \citenamefont {{Gonz{\'a}lez-Tudela}},\ and\
  \citenamefont {Cirac}}]{wei_generation_2021}%
  \BibitemOpen
  \bibfield  {author} {\bibinfo {author} {\bibfnamefont {Z.-Y.}\ \bibnamefont
  {Wei}}, \bibinfo {author} {\bibfnamefont {D.}~\bibnamefont {Malz}}, \bibinfo
  {author} {\bibfnamefont {A.}~\bibnamefont {{Gonz{\'a}lez-Tudela}}}, \ and\
  \bibinfo {author} {\bibfnamefont {J.~I.}\ \bibnamefont {Cirac}},\ }\bibinfo
  {title} {\emph {Generation of {{Photonic Matrix Product States}} with
  {{Rydberg Atomic Arrays}}}},\ \href {\doibase
  10.1103/PhysRevResearch.3.023021} {\bibfield  {journal} {\bibinfo  {journal}
  {Physical Review Research}\ }\textbf {\bibinfo {volume} {3}},\ \bibinfo
  {pages} {023021} (\bibinfo {year} {2021})},\ \Eprint
  {http://arxiv.org/abs/2011.03919} {arXiv:2011.03919}\BibitemShut {NoStop}%
\bibitem [{\citenamefont {Honer}\ \emph {et~al.}(2011)\citenamefont {Honer},
  \citenamefont {L{\"o}w}, \citenamefont {Weimer}, \citenamefont {Pfau},\ and\
  \citenamefont {B{\"u}chler}}]{honer_artificial_2011}%
  \BibitemOpen
  \bibfield  {author} {\bibinfo {author} {\bibfnamefont {J.}~\bibnamefont
  {Honer}}, \bibinfo {author} {\bibfnamefont {R.}~\bibnamefont {L{\"o}w}},
  \bibinfo {author} {\bibfnamefont {H.}~\bibnamefont {Weimer}}, \bibinfo
  {author} {\bibfnamefont {T.}~\bibnamefont {Pfau}}, \ and\ \bibinfo {author}
  {\bibfnamefont {H.~P.}\ \bibnamefont {B{\"u}chler}},\ }\bibinfo {title}
  {\emph {Artificial Atoms Can Do More than Atoms: {{Deterministic}} Single
  Photon Subtraction from Arbitrary Light Fields}},\ \href {\doibase
  10.1103/PhysRevLett.107.093601} {\bibfield  {journal} {\bibinfo  {journal}
  {Physical Review Letters}\ }\textbf {\bibinfo {volume} {107}},\ \bibinfo
  {pages} {093601} (\bibinfo {year} {2011})},\ \Eprint
  {http://arxiv.org/abs/1103.1319} {arXiv:1103.1319}\BibitemShut {NoStop}%
\bibitem [{\citenamefont {Murray}\ \emph {et~al.}(2018)\citenamefont {Murray},
  \citenamefont {Mirgorodskiy}, \citenamefont {Tresp}, \citenamefont {Braun},
  \citenamefont {{Paris-Mandoki}}, \citenamefont {Gorshkov}, \citenamefont
  {Hofferberth},\ and\ \citenamefont {Pohl}}]{murray_photon_2018}%
  \BibitemOpen
  \bibfield  {author} {\bibinfo {author} {\bibfnamefont {C.~R.}\ \bibnamefont
  {Murray}}, \bibinfo {author} {\bibfnamefont {I.}~\bibnamefont
  {Mirgorodskiy}}, \bibinfo {author} {\bibfnamefont {C.}~\bibnamefont {Tresp}},
  \bibinfo {author} {\bibfnamefont {C.}~\bibnamefont {Braun}}, \bibinfo
  {author} {\bibfnamefont {A.}~\bibnamefont {{Paris-Mandoki}}}, \bibinfo
  {author} {\bibfnamefont {A.~V.}\ \bibnamefont {Gorshkov}}, \bibinfo {author}
  {\bibfnamefont {S.}~\bibnamefont {Hofferberth}}, \ and\ \bibinfo {author}
  {\bibfnamefont {T.}~\bibnamefont {Pohl}},\ }\bibinfo {title} {\emph {Photon
  {{Subtraction}} by {{Many-Body Decoherence}}}},\ \href {\doibase
  10.1103/PhysRevLett.120.113601} {\bibfield  {journal} {\bibinfo  {journal}
  {Physical Review Letters}\ }\textbf {\bibinfo {volume} {120}},\ \bibinfo
  {pages} {113601} (\bibinfo {year} {2018})}\BibitemShut {NoStop}%
\bibitem [{\citenamefont {Burgers}\ \emph {et~al.}(2022)\citenamefont
  {Burgers}, \citenamefont {Ma}, \citenamefont {Saskin}, \citenamefont
  {Wilson}, \citenamefont {Alarc{\'o}n}, \citenamefont {Greene},\ and\
  \citenamefont {Thompson}}]{burgers_controlling_2022}%
  \BibitemOpen
  \bibfield  {author} {\bibinfo {author} {\bibfnamefont {A.~P.}\ \bibnamefont
  {Burgers}}, \bibinfo {author} {\bibfnamefont {S.}~\bibnamefont {Ma}},
  \bibinfo {author} {\bibfnamefont {S.}~\bibnamefont {Saskin}}, \bibinfo
  {author} {\bibfnamefont {J.}~\bibnamefont {Wilson}}, \bibinfo {author}
  {\bibfnamefont {M.~A.}\ \bibnamefont {Alarc{\'o}n}}, \bibinfo {author}
  {\bibfnamefont {C.~H.}\ \bibnamefont {Greene}}, \ and\ \bibinfo {author}
  {\bibfnamefont {J.~D.}\ \bibnamefont {Thompson}},\ }\bibinfo {title} {\emph
  {Controlling {{Rydberg Excitations Using Ion-Core Transitions}} in
  {{Alkaline-Earth Atom-Tweezer Arrays}}}},\ \href {\doibase
  10.1103/PRXQuantum.3.020326} {\bibfield  {journal} {\bibinfo  {journal} {PRX
  Quantum}\ }\textbf {\bibinfo {volume} {3}},\ \bibinfo {pages} {020326}
  (\bibinfo {year} {2022})}\BibitemShut {NoStop}%
\bibitem [{\citenamefont {Covey}\ \emph {et~al.}(2019)\citenamefont {Covey},
  \citenamefont {Sipahigil}, \citenamefont {Szoke}, \citenamefont {Sinclair},
  \citenamefont {Endres},\ and\ \citenamefont
  {Painter}}]{covey_telecom-band_2019}%
  \BibitemOpen
  \bibfield  {author} {\bibinfo {author} {\bibfnamefont {J.~P.}\ \bibnamefont
  {Covey}}, \bibinfo {author} {\bibfnamefont {A.}~\bibnamefont {Sipahigil}},
  \bibinfo {author} {\bibfnamefont {S.}~\bibnamefont {Szoke}}, \bibinfo
  {author} {\bibfnamefont {N.}~\bibnamefont {Sinclair}}, \bibinfo {author}
  {\bibfnamefont {M.}~\bibnamefont {Endres}}, \ and\ \bibinfo {author}
  {\bibfnamefont {O.}~\bibnamefont {Painter}},\ }\bibinfo {title} {\emph
  {Telecom-{{Band Quantum Optics}} with {{Ytterbium Atoms}} and {{Silicon
  Nanophotonics}}}},\ \href {\doibase 10.1103/PhysRevApplied.11.034044}
  {\bibfield  {journal} {\bibinfo  {journal} {Physical Review Applied}\
  }\textbf {\bibinfo {volume} {11}},\ \bibinfo {pages} {034044} (\bibinfo
  {year} {2019})}\BibitemShut {NoStop}%
\bibitem [{\citenamefont {Wilson}\ \emph {et~al.}(2022)\citenamefont {Wilson},
  \citenamefont {Saskin}, \citenamefont {Meng}, \citenamefont {Ma},
  \citenamefont {Dilip}, \citenamefont {Burgers},\ and\ \citenamefont
  {Thompson}}]{wilson_trapping_2022}%
  \BibitemOpen
  \bibfield  {author} {\bibinfo {author} {\bibfnamefont {J.~T.}\ \bibnamefont
  {Wilson}}, \bibinfo {author} {\bibfnamefont {S.}~\bibnamefont {Saskin}},
  \bibinfo {author} {\bibfnamefont {Y.}~\bibnamefont {Meng}}, \bibinfo {author}
  {\bibfnamefont {S.}~\bibnamefont {Ma}}, \bibinfo {author} {\bibfnamefont
  {R.}~\bibnamefont {Dilip}}, \bibinfo {author} {\bibfnamefont {A.~P.}\
  \bibnamefont {Burgers}}, \ and\ \bibinfo {author} {\bibfnamefont {J.~D.}\
  \bibnamefont {Thompson}},\ }\bibinfo {title} {\emph {Trapping {{Alkaline
  Earth Rydberg Atoms Optical Tweezer Arrays}}}},\ \href {\doibase
  10.1103/PhysRevLett.128.033201} {\bibfield  {journal} {\bibinfo  {journal}
  {Physical Review Letters}\ }\textbf {\bibinfo {volume} {128}},\ \bibinfo
  {pages} {033201} (\bibinfo {year} {2022})}\BibitemShut {NoStop}%
\bibitem [{\citenamefont {{Asenjo-Garcia}}\ \emph {et~al.}(2019)\citenamefont
  {{Asenjo-Garcia}}, \citenamefont {Kimble},\ and\ \citenamefont
  {Chang}}]{asenjo-garcia_optical_2019}%
  \BibitemOpen
  \bibfield  {author} {\bibinfo {author} {\bibfnamefont {A.}~\bibnamefont
  {{Asenjo-Garcia}}}, \bibinfo {author} {\bibfnamefont {H.~J.}\ \bibnamefont
  {Kimble}}, \ and\ \bibinfo {author} {\bibfnamefont {D.~E.}\ \bibnamefont
  {Chang}},\ }\bibinfo {title} {\emph {Optical Waveguiding by Atomic
  Entanglement in Multilevel Atom Arrays}},\ \href {\doibase
  10.1073/pnas.1911467116} {\bibfield  {journal} {\bibinfo  {journal}
  {Proceedings of the National Academy of Sciences}\ }\textbf {\bibinfo
  {volume} {116}},\ \bibinfo {pages} {25503} (\bibinfo {year}
  {2019})}\BibitemShut {NoStop}%
\bibitem [{\citenamefont {Gorshkov}\ \emph {et~al.}(2007)\citenamefont
  {Gorshkov}, \citenamefont {Andre}, \citenamefont {Lukin},\ and\ \citenamefont
  {Sorensen}}]{gorshkov_photon_2007}%
  \BibitemOpen
  \bibfield  {author} {\bibinfo {author} {\bibfnamefont {A.~V.}\ \bibnamefont
  {Gorshkov}}, \bibinfo {author} {\bibfnamefont {A.}~\bibnamefont {Andre}},
  \bibinfo {author} {\bibfnamefont {M.~D.}\ \bibnamefont {Lukin}}, \ and\
  \bibinfo {author} {\bibfnamefont {A.~S.}\ \bibnamefont {Sorensen}},\
  }\bibinfo {title} {\emph {Photon Storage in {{Lambda-type}} Optically Dense
  Atomic Media. {{II}}. {{Free-space}} Model}},\ \href {\doibase
  10.1103/PhysRevA.76.033805} {\bibfield  {journal} {\bibinfo  {journal}
  {Physical Review A}\ }\textbf {\bibinfo {volume} {76}},\ \bibinfo {pages}
  {033805} (\bibinfo {year} {2007})},\ \Eprint
  {http://arxiv.org/abs/quant-ph/0612083} {arXiv:quant-ph/0612083}\BibitemShut
  {NoStop}%
\bibitem [{\citenamefont {Wallentowitz}\ and\ \citenamefont
  {Vogel}(1996)}]{wallentowitz_unbalanced_1996}%
  \BibitemOpen
  \bibfield  {author} {\bibinfo {author} {\bibfnamefont {S.}~\bibnamefont
  {Wallentowitz}}\ and\ \bibinfo {author} {\bibfnamefont {W.}~\bibnamefont
  {Vogel}},\ }\bibinfo {title} {\emph {Unbalanced Homodyning for Quantum State
  Measurements}},\ \href {\doibase 10.1103/PhysRevA.53.4528} {\bibfield
  {journal} {\bibinfo  {journal} {Physical Review A}\ }\textbf {\bibinfo
  {volume} {53}},\ \bibinfo {pages} {4528} (\bibinfo {year}
  {1996})}\BibitemShut {NoStop}%
\bibitem [{\citenamefont {Banaszek}\ and\ \citenamefont
  {W{\'o}dkiewicz}(1996)}]{banaszek_direct_1996}%
  \BibitemOpen
  \bibfield  {author} {\bibinfo {author} {\bibfnamefont {K.}~\bibnamefont
  {Banaszek}}\ and\ \bibinfo {author} {\bibfnamefont {K.}~\bibnamefont
  {W{\'o}dkiewicz}},\ }\bibinfo {title} {\emph {Direct {{Probing}} of {{Quantum
  Phase Space}} by {{Photon Counting}}}},\ \href {\doibase
  10.1103/PhysRevLett.76.4344} {\bibfield  {journal} {\bibinfo  {journal}
  {Physical Review Letters}\ }\textbf {\bibinfo {volume} {76}},\ \bibinfo
  {pages} {4344} (\bibinfo {year} {1996})}\BibitemShut {NoStop}%
\bibitem [{\citenamefont {Nehra}\ \emph {et~al.}(2019)\citenamefont {Nehra},
  \citenamefont {Win}, \citenamefont {Eaton}, \citenamefont {Shahrokhshahi},
  \citenamefont {Sridhar}, \citenamefont {Gerrits}, \citenamefont {Lita},
  \citenamefont {Nam},\ and\ \citenamefont
  {Pfister}}]{nehra_state-independent_2019}%
  \BibitemOpen
  \bibfield  {author} {\bibinfo {author} {\bibfnamefont {R.}~\bibnamefont
  {Nehra}}, \bibinfo {author} {\bibfnamefont {A.}~\bibnamefont {Win}}, \bibinfo
  {author} {\bibfnamefont {M.}~\bibnamefont {Eaton}}, \bibinfo {author}
  {\bibfnamefont {R.}~\bibnamefont {Shahrokhshahi}}, \bibinfo {author}
  {\bibfnamefont {N.}~\bibnamefont {Sridhar}}, \bibinfo {author} {\bibfnamefont
  {T.}~\bibnamefont {Gerrits}}, \bibinfo {author} {\bibfnamefont
  {A.}~\bibnamefont {Lita}}, \bibinfo {author} {\bibfnamefont {S.~W.}\
  \bibnamefont {Nam}}, \ and\ \bibinfo {author} {\bibfnamefont
  {O.}~\bibnamefont {Pfister}},\ }\bibinfo {title} {\emph {State-Independent
  Quantum State Tomography by Photon-Number-Resolving Measurements}},\ \href
  {\doibase 10.1364/OPTICA.6.001356} {\bibfield  {journal} {\bibinfo  {journal}
  {Optica}\ }\textbf {\bibinfo {volume} {6}},\ \bibinfo {pages} {1356}
  (\bibinfo {year} {2019})}\BibitemShut {NoStop}%
\bibitem [{\citenamefont {{Foss-Feig}}\ \emph {et~al.}(2017)\citenamefont
  {{Foss-Feig}}, \citenamefont {Young}, \citenamefont {Albert}, \citenamefont
  {Gorshkov},\ and\ \citenamefont {Maghrebi}}]{foss-feig_solvable_2017}%
  \BibitemOpen
  \bibfield  {author} {\bibinfo {author} {\bibfnamefont {M.}~\bibnamefont
  {{Foss-Feig}}}, \bibinfo {author} {\bibfnamefont {J.~T.}\ \bibnamefont
  {Young}}, \bibinfo {author} {\bibfnamefont {V.~V.}\ \bibnamefont {Albert}},
  \bibinfo {author} {\bibfnamefont {A.~V.}\ \bibnamefont {Gorshkov}}, \ and\
  \bibinfo {author} {\bibfnamefont {M.~F.}\ \bibnamefont {Maghrebi}},\
  }\bibinfo {title} {\emph {Solvable {{Family}} of {{Driven-Dissipative
  Many-Body Systems}}}},\ \href {\doibase 10.1103/PhysRevLett.119.190402}
  {\bibfield  {journal} {\bibinfo  {journal} {Physical Review Letters}\
  }\textbf {\bibinfo {volume} {119}},\ \bibinfo {pages} {190402} (\bibinfo
  {year} {2017})}\BibitemShut {NoStop}%
\end{thebibliography}%

\clearpage
\raggedbottom
\appendix
\section{Efficient Numerical Algorithm}\label{ap:numerics}
In this section, we present a construction of the truncated basis which can be used to efficiently solve the many-body problem defined by the master equation:
\begin{align}
    \dot\rho(t)=-i[\hat H,\rho]+\gamma\left(\sum_i \hat n_i\rho \hat n_i-\frac{1}{2}\{\hat n_i,\rho\}\right),
\end{align}
where $\hat H=\Omega\sum_{i=1}^N\hat\sigma_{rs}^{(i)}+\text{h.c.}$ and $\hat n_i:=\ketbra{r_i}{r_i}$. In order to solve this equation numerically, we want to write down the Lindbladian 
$
\mathcal{L}[\cdot]=-i[\hat H,\cdot]+\gamma\left(\sum_i \hat n_i\cdot \hat n_i-\frac{1}{2}\{\hat n_i,\cdot\}\right)
$ 
as a superoperator $\mathcal{L}[\cdot]=: i\mathcal{H}[\cdot]+\gamma \mathcal{D}[\cdot]$ in an efficiently constructed basis. We will find that the size of this basis grows linearly in $n$, which we take to be fixed. We use the notation of \cite{foss-feig_solvable_2017}, where density matrices are written as superkets with double brackets $|\rho\rrangle = \sum_{i,j}\rho_{ij}\ketbra{i}{j}$, superbras are defined as $\llangle \rho|=|\rho\rrangle^\dagger$, and the inner product is given by $\llangle \mu|\nu\rrangle=\Tr[\mu^\dagger\nu]$.

We want to find the matrix forms $\mathcal{H}$  and $\mathcal{D}$ of our superoperators, where $\mathcal{H}_{\mu\nu}:= \llangle \mu|\mathcal{H}|\nu\rrangle$ and similarly for $\mathcal{D}_{\mu\nu}$. To do this, we will exploit the symmetry of the Lindbladian. The initial state $\ketbra{S}{S}$ is symmetric under permutation of the atomic sites, and $\mathcal{L}$ respects this symmetry, so we can restrict our basis to only permutation-symmetric states. We can further recognize that both the initial state and the Lindbladian are symmetric with respect to $\mathcal{O}_{gg}:=\sum_{i=1}^N\hat\sigma_{gg}^{(i)}$, in the sense that $\mathcal{O}_{gg}\ketbra{S}{S}=\ketbra{S}{S}$ and $[\mathcal{O}_{gg},\hat H]=[\mathcal{O}_{gg},\hat n_i]=0$. This allows us to block diagonalize $\mathcal{L}$ in terms of the $\hat\sigma_{gg}$ population:

\begin{figure*}
    \centering
    \includegraphics[width = 0.99\textwidth]{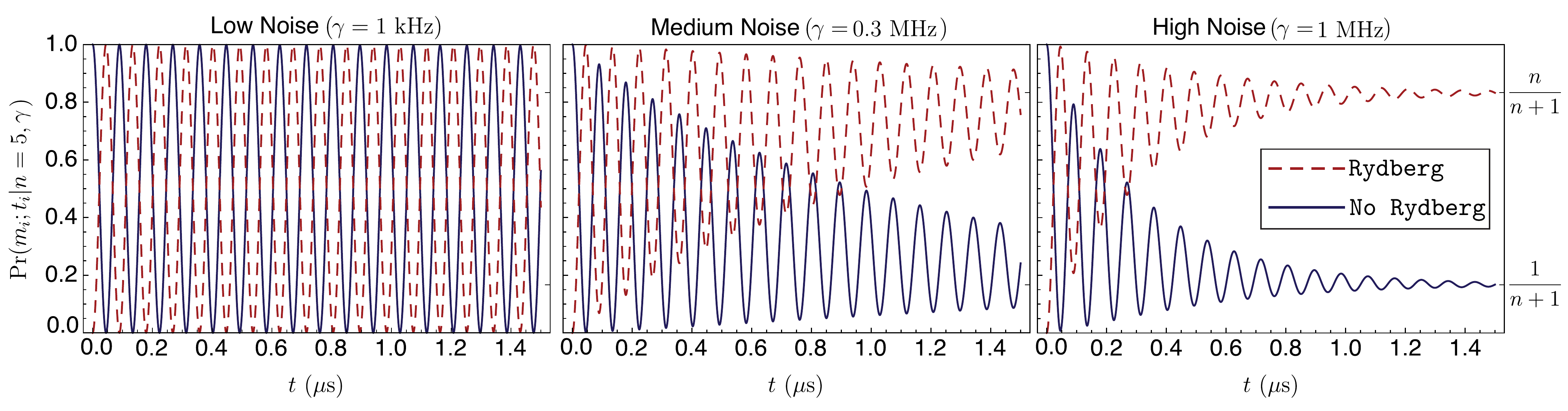}
    \caption{Rydberg oscillation dynamics in the presence of noise with $n=5$, $N=10$, and $\Omega=2\pi\times 2.5$ MHz. Notice that the populations approach the steady-state values of $\frac{n}{n+1}=5/6$ and $\frac{1}{n+1}=1/6$.}
    \label{fig:efficient}
\end{figure*}

\begin{widetext}
    \begin{align}
    \begin{split}
        |\rho^{ss}_j\rrangle &= \mathcal{N}^{ss}_j \sum_\text{perms}\left(\hat\sigma_{ss}\right)^{\otimes n-j}\left(\hat\sigma_{sg}\otimes\hat\sigma_{gs}\right)^{\otimes j}\left(\hat\sigma_{gg}\right)^{\otimes N-n-j},\\
        |\rho^{rs}_j\rrangle &= \mathcal{N}^{rs}_j \sum_\text{perms}\hat\sigma_{rs}\otimes\left(\hat\sigma_{sg}\otimes\hat\sigma_{gs}\right)^{\otimes j}\left(\hat\sigma_{ss}\right)^{\otimes n-j-1}\left(\hat\sigma_{gg}\right)^{\otimes N-n-j},\\
        |\rho^{sr}_j\rrangle &= |\rho^{rs}_j\rrangle^\dagger,\\
        |\rho^{rg}_j\rrangle &= \mathcal{N}^{rg}_j \sum_\text{perms}\hat\sigma_{rg}\otimes\left(\hat\sigma_{sg}\right)^{\otimes j-1}\left(\hat\sigma_{gs}\right)^{\otimes j}\left(\hat\sigma_{ss}\right)^{\otimes n-j}\left(\hat\sigma_{gg}\right)^{\otimes N-n-j},\\
        |\rho^{gr}_j\rrangle &= |\rho^{rg}_j\rrangle^\dagger,\\
        |\rho^{rr}_j\rrangle &= \mathcal{N}^{rr}_j \sum_\text{perms}\hat\sigma_{rr}\otimes\left(\hat\sigma_{sg}\otimes\hat\sigma_{gs}\right)^{\otimes j}\left(\hat\sigma_{ss}\right)^{\otimes n-j-1}\left(\hat\sigma_{gg}\right)^{\otimes N-n-j},\\
        |\rho^{rs,gr}_j\rrangle &= \mathcal{N}^{rs,gr}_j \sum_\text{perms}\hat\sigma_{rs}\otimes\hat\sigma_{gr}\otimes\left(\hat\sigma_{sg}\left)^{\otimes j}\right(\hat\sigma_{gs}\right)^{\otimes j-1}\left(\hat\sigma_{ss}\right)^{\otimes n-j-1}\left(\hat\sigma_{gg}\right)^{\otimes N-n-j},\\
        |\rho^{sr,rg}_j\rrangle &=|\rho^{rs,gr}_j\rrangle^\dagger,\\
        |\rho^{rs,sr}_j\rrangle &= \mathcal{N}^{rs,sr}_j \sum_\text{perms}\hat\sigma_{rs}\otimes\hat\sigma_{sr}\otimes\left(\hat\sigma_{sg}\otimes\hat\sigma_{gs}\right)^{\otimes j}\left(\hat\sigma_{ss}\right)^{\otimes n-j-2}\left(\hat\sigma_{gg}\right)^{\otimes N-n-j},\\
        |\rho^{rg,gr}_j \rrangle &= \mathcal{N}^{rg,gr}_j \sum_\text{perms}\hat\sigma_{rg}\otimes\hat\sigma_{gr}\otimes\left(\hat\sigma_{sg}\otimes\hat\sigma_{gs}\right)^{\otimes j-1}\left(\hat\sigma_{ss}\right)^{\otimes n-j}\left(\hat\sigma_{gg}\right)^{\otimes N-n-j},\\
    \end{split}
    \end{align}
\end{widetext}
where the sum is over permutations of the on-site density matrices across sites and $\hat \sigma_{\mu\nu}:=\ketbra{\mu}{\nu}$. $j=0,\ldots,n$ indexes the blocks of the Hamiltonian superoperator $\mathcal{H}=\bigoplus_{j=0}^n \mathcal{H}^{(j)}$ and the $\mathcal{N}$ are normalization constants such that $\llangle\rho_i|\rho_i\rrangle=1$. This basis also diagonalizes $\mathcal{D}_{\mu\nu}$ entirely.

We can explicitly express $\mathcal{H}_{\mu\nu}^{(j)}$ in terms of the normalization coefficients of the basis states:
\begin{widetext}
\begin{align}
   \mathcal{H}^{(j)} = \Omega_n\left(
\begin{smallmatrix}
 0 & -\frac{\mathcal{N}^{ss}_j}{\mathcal{N}^{rs}_j} & \frac{\mathcal{N}^{ss}_j}{\mathcal{N}^{sr}_j} & -\frac{\mathcal{N}^{ss}_j}{\mathcal{N}^{rg}_j} & \frac{\mathcal{N}^{ss}_j}{\mathcal{N}^{gr}_j} & 0 & 0 & 0 & 0 & 0 \\
 \frac{-(n-j) \mathcal{N}^{rs}_j}{\mathcal{N}^{ss}_j} & 0 & 0 & 0 & 0 & \frac{\mathcal{N}^{rs}_j}{\mathcal{N}^{rr}_j} & \frac{\mathcal{N}^{rs}_j}{\mathcal{N}^{rs,gr}_j} & 0 & \frac{\mathcal{N}^{rs}_j}{\mathcal{N}^{rs,sr}_j} & 0 \\
 \frac{(n-j) \mathcal{N}^{sr}_j}{\mathcal{N}^{ss}_j} & 0 & 0 & 0 & 0 & -\frac{\mathcal{N}^{sr}_j}{\mathcal{N}^{rr}_j} & 0 & -\frac{\mathcal{N}^{sr}_j}{\mathcal{N}^{sr,rg}_j} & -\frac{\mathcal{N}^{sr}_j}{\mathcal{N}^{rs,sr}_j} & 0 \\
 -\frac{j \mathcal{N}^{rg}_j}{\mathcal{N}^{ss}_j} & 0 & 0 & 0 & 0 & 0 & 0 & \frac{\mathcal{N}^{rg}_j}{\mathcal{N}^{sr,rg}_j} & 0 & \frac{\mathcal{N}^{rg}_j}{\mathcal{N}^{rg,gr}_j} \\
 \frac{j \mathcal{N}^{gr}_j}{\mathcal{N}^{ss}_j} & 0 & 0 & 0 & 0 & 0 & -\frac{\mathcal{N}^{gr}_j}{\mathcal{N}^{rs,gr}_j} & 0 & 0 & -\frac{\mathcal{N}^{gr}_j}{\mathcal{N}^{rg,gr}_j} \\
 0 & \frac{\mathcal{N}^{rr}_j}{\mathcal{N}^{rs}_j} & -\frac{\mathcal{N}^{rr}_j}{\mathcal{N}^{sr}_j} & 0 & 0 & 0 & 0 & 0 & 0 & 0 \\
 0 & \frac{j \mathcal{N}^{rs,gr}_j}{\mathcal{N}^{rs}_j} & 0 & 0 & \frac{(j-n) \mathcal{N}^{rs,gr}_j}{\mathcal{N}^{gr}_j} & 0 & 0 & 0 & 0 & 0 \\
 0 & 0 & -\frac{j \mathcal{N}^{sr,rg}_j}{\mathcal{N}^{sr}_j} & \frac{(n-j) \mathcal{N}^{sr,rg}_j}{\mathcal{N}^{rg}_j} & 0 & 0 & 0 & 0 & 0 & 0 \\
 0 & \frac{(n-j-1) \mathcal{N}^{rs,sr}_j}{\mathcal{N}^{rs}_j} & \frac{-(n-j-1) \mathcal{N}^{rs,sr}_j}{\mathcal{N}^{sr}_j} & 0 & 0 & 0 & 0 & 0 & 0 & 0 \\
 0 & 0 & 0 & \frac{j \mathcal{N}^{rg,gr}_j}{\mathcal{N}^{rg}_j} & -\frac{j \mathcal{N}^{rg,gr}_j}{\mathcal{N}^{gr}_j} & 0 & 0 & 0 & 0 & 0 \\
\end{smallmatrix}
\right)
\end{align}
\end{widetext}

Because the truncated basis diagonalizes $\mathcal{D}$, we have a particularly simple matrix representation of $\mathcal{D}^{(j)}$:
\begin{align}
    \mathcal{D}^{(j)}=-\gamma\left(
\begin{array}{cccccccccc}
 0 & 0 & 0 & 0 & 0 & 0 & 0 & 0 & 0 & 0 \\
 0 & \frac{1}{2} & 0 & 0 & 0 & 0 & 0 & 0 & 0 & 0 \\
 0 & 0 & \frac{1}{2} & 0 & 0 & 0 & 0 & 0 & 0 & 0 \\
 0 & 0 & 0 & \frac{1}{2} & 0 & 0 & 0 & 0 & 0 & 0 \\
 0 & 0 & 0 & 0 & \frac{1}{2} & 0 & 0 & 0 & 0 & 0 \\
 0 & 0 & 0 & 0 & 0 & 0 & 0 & 0 & 0 & 0 \\
 0 & 0 & 0 & 0 & 0 & 0 & 1 & 0 & 0 & 0 \\
 0 & 0 & 0 & 0 & 0 & 0 & 0 & 1 & 0 & 0 \\
 0 & 0 & 0 & 0 & 0 & 0 & 0 & 0 & 1 & 0 \\
 0 & 0 & 0 & 0 & 0 & 0 & 0 & 0 & 0 & 1 \\
\end{array}
\right).
\end{align}

Notice that the only populations in $\rho$ are given by $|\rho^{ss}_0\rrangle$ and $|\rho^{rr}_0\rrangle$. It is therefore only necessary to solve for the dynamics in the $j=0$ block of $\rho$ in order to calculate $\Pr(M_T|n,\gamma)$. We numerically solve the master equation for $n=5$ and $N=10$ and a variety of dephasing rates $\gamma$, the results of which are displayed in \cref{fig:efficient}.

\section{Generalization to Mixed States}\label{ap:mixed}
In this section, we continue the discussion of the generalization of the protocol to mixed initial photonic states. It is straightforward to see that the measurement dynamics of the protocol are insensitive to coherences between states of different photon number. Consider the evolution of the density matrix of the system $\rho$ under the unitary dynamics generated by $\hat H_\text{coll}$, governed by the master equation $\dot\rho(t)=-i[\hat H_\text{coll},\rho]$. Clearly these dynamics do not couple states with different photon numbers, which implies the insensitivity to coherences of the dynamics of the populations. This means that the populations of two initial density matrices with the same populations and different coherences will evolve in the same way under the protocol. In particular, this implies that the populations of any density matrix will evolve in the same way as the diagonal density matrix with the same populations. This has two important implications. 

First, it explains why it is not possible to choose drive times $\tau_i$ which steer the state of the system toward a particular photon number state. Because the evolution of the populations is indistinguishable from that of a classical mixture $\rho_\text{class}=\sum_n p_n \ketbra{n}{n}$, the dynamics serve to distill the state $\ket{n}$ with probability $p_n=|c_n|^2$, regardless of the measurement pattern chosen. However, different measurement patterns may lead to faster convergence to $\ket{n}$, as is discussed \cref{ap:adaptive}. 

Second, it implies that the path to convergence recorded in the measurement record does not directly encode any information beyond the final value to which the protocol converged. This is because the populations evolve as if the density matrix were a classical mixture, so the measurement dynamics are consistent with a single $\ket{n}$ being sampled from this distribution at the beginning of the protocol. However, it is still possible to learn something about the initial populations of other number states via the prior. As an example, consider a uniform prior taken over $P_1=(0,0.9,0.1)$ and $P_2=(0,0.1,0.9)$. If an experiment converges to $n=1$, we know that $P_1$ was more likely and therefore it is likely that there was little initial population in $n=2$, even though all we directly measured was that there was initially some population in $n=1$.

\section{Adaptive Optimization Strategy}\label{ap:adaptive}
In this section, we illustrate through a minimal example that optimizing the drives times $\{\tau_i\}$ is exponentially difficult. In order to precisely define the optimal strategy for choosing drive times $\{\tau_i\}$, it is necessary to first fix a figure of merit. We here consider the accuracy of the initial state inference as given by the maximum likelihood estimate (MLE) generated by the measurement record. The expected value of the MLE over experimental realizations is called the \textit{fidelity} and denoted by $\mathcal{F}$. More explicitly, for a set of potential initial photon number distributions $\{P_\alpha\}_N$ and $T$ observation cycles, we have
\begin{align}
    \mathcal{F}_T=\sum_{M_T}\underset{P_\alpha}{\text{max}}\text{ Pr}(M_T|P_\alpha)\text{Pr}(P_\alpha).
\end{align}

It is natural to consider whether a local-in-time strategy---that is, choosing each $\tau_i$ independently to maximize each $\mathcal{F}_i$---might yield a globally optimal set of times $\{\tau_i\}_{i=1,\ldots,T}$ with respect to the fidelity after some fixed number $T$ of observation cycles, $\mathcal{F}_T$. This task is appropriate for the regime in which the measurement time is large compared to the standard deviation of the oscillation time, so that a fixed number of observation cycles approximately corresponds to fixed total time. We have demonstrated numerically that such a local strategy is not in general optimal. 

In a toy state discrimination task wherein the protocol had two observation cycles ($T=2$) to distinguish between two potential initial states $P_1=(0,\frac{1}{3},\frac{1}{3},0,\frac{1}{3},0)$ and $P_2=(0,0,0,1,0,0)$, the local strategy led after the first cycle---$\mathcal{F}_1^\text{local}=96.59\%$, $\mathcal{F}_1^\text{global}=96.52\%$---but was overtaken in the second cycle---$\mathcal{F}_1^\text{local}=99.84\%$, $\mathcal{F}_1^\text{global}=99.89\%$.

This demonstrates that allowing for $\tau_{i}$ which yield suboptimal values of intermediate $\mathcal{F}_i$ avails measurement patterns with higher values of $\mathcal{F}_T$ at the conclusion of the protocol, so that the local-in-time strategy is not optimal. However, these differences can be quite small, as in this toy example.

For a set number of observation cycles $T$, the optimal strategy---that which maximizes $\mathcal F_T$---can still in principle be evaluated numerically. However, the absence of an optimal local-in-time strategy implies that the cost of finding the globally optimal strategy grows exponentially in the number of observation cycles $T$.
\end{document}